\documentclass[12pt]{article}
\usepackage{amsmath,amsfonts,feynmp,epsf}
\usepackage{amssymb}
\usepackage{graphicx}
\usepackage{grffile}
\input epsf
\usepackage[nosort]{cite}
\usepackage[english]{babel}

\makeatletter\@addtoreset{equation}{section}\makeatother

{\catcode`\|=\active
  \gdef\Braket#1{\left<\mathcode`\|"8000\let|\BraVert {#1}\right>}}
\def\BraVert{\egroup\,\mid@vertical\,\bgroup}
{\catcode`\|=\active
  \gdef\set#1{\mathinner{\lbrace\,{\mathcode`\|"8000\let|\midvert #1}\,\rbrace}}
  \gdef\Set#1{\left\{\:{\mathcode`\|"8000\let|\SetVert #1}\:\right\}}}
\def\midvert{\egroup\mid\bgroup}
\def\SetVert{\egroup\;\mid@vertical\;\bgroup}

\def\be{\begin{equation}}
\def\ee{\end{equation}}
\def\bea{\begin{eqnarray}}
\def\eea{\end{eqnarray}}
\def\ie{\begin{equation}\begin{aligned}}
\def\fe{\end{aligned}\end{equation}}

\makeatletter\@addtoreset{equation}{section}\makeatother

\newcommand{\preprint}[1]{\begin{table}[t]  
             \begin{flushright}               
             {#1}                             
             \end{flushright}                 
             \end{table}}                     
\renewcommand{\title}[1]{\vbox{\center\LARGE{#1}}\vspace{5mm}}
\renewcommand{\author}[1]{\vbox{\center#1}\vspace{5mm}}
\newcommand{\address}[1]{\vbox{\center\em#1}}
\newcommand{\email}[1]{\vbox{\center\tt#1}\vspace{5mm}}

\newcommand{\A}{{\alpha}}
\newcommand{\B}{{\beta}}
\newcommand{\C}{{\gamma}}
\newcommand{\D}{{\delta}}
\newcommand{\da}{{\dot\alpha}}
\newcommand{\db}{{\dot\beta}}
\newcommand{\dc}{{\dot\gamma}}
\newcommand{\dd}{{\dot\delta}}

\begin{document}
\begin{titlepage}
\preprint{pi-qf\&strings-292}
\begin{center}
\vskip 1cm

\title{The Higher Spin/Vector Model Duality}

\author{Simone Giombi$^{1,a}$ and
Xi Yin$^{2,b}$}

\address{${}^1$Perimeter Institute for Theoretical Physics, Waterloo, Ontario, N2L 2Y5, Canada}
\address{
${}^2$Center for the Fundamental Laws of Nature,
Jefferson Physical Laboratory,\\
Harvard University,
Cambridge, MA 02138 USA}

\email{$^a$sgiombi@pitp.ca,
$^b$xiyin@fas.harvard.edu}

\end{center}

\abstract{This paper is mainly a review of the dualities between Vasiliev's higher spin gauge theories in $AdS_4$ and three dimensional large $N$ vector models, with focus on the holographic calculation of correlation functions of higher spin currents. We also present some new results in the computation of parity odd structures in the three point functions in parity violating Vasiliev theories. }

\vfill

\end{titlepage}

\eject \tableofcontents

\section{Introduction} 

The holographic duality between Vasiliev's higher spin gauge theory in $AdS_4$ and $O(N)$ vector models was conjectured a decade ago \cite{Klebanov:2002ja}\cite{Sezgin:2003pt} (see also \cite{Sundborg:1999ue, HaggiMani:2000ru, Konstein:2000bi, Shaynkman:2001ip, Sezgin:2001zs, Vasiliev:2001zy, Mikhailov:2002bp} and in particular \cite{Sezgin:2002rt} for earlier closely related work). The higher spin/vector model duality could be regarded as the simplest nontrivial examples of AdS/CFT correspondence. In particular, the spectrum of operators in the CFT is not renormalized at infinite $N$ \cite{Giombi:2011kc}, and correspondingly the spectrum of fields in the bulk theory is simple. The nontrivial content of the duality, at least in perturbation theory, therefore lies in the agreement between correlation functions in the bulk higher spin gauge theory and the boundary vector model CFT. Due to technical complications, correlators in Vasiliev theory resisted attempts of direct calculation until \cite{Giombi:2009wh, Giombi:2010vg}, where the three point functions of higher spin currents are computed and shown to match with those of the free and critical $O(N)$ vector models.  Substantial progress in higher spin holography has been made in the last few years, including generalizing the conjecture to parity violating Vasiliev theories and Chern-Simons vector models \cite{Giombi:2011kc, Chang:2012kt}, structure of correlation functions \cite{Giombi:2011rz, Costa:2011mg, Zhiboedov:2012bm}, proof of CFT/higher spin version of Coleman-Mandula theorem \cite{Maldacena:2011jn, Maldacena:2012sf}, exact large $N$ computations in Chern-Simons vector models \cite{Giombi:2011kc, OferUpcoming, Jain:2012qi}, supersymmetric extension and symmetry breaking, and connection between Vasiliev's higher spin gauge theory and string theory \cite{Chang:2012kt}. Some approaches towards deriving the higher spin/vector model duality from first principles were investigated in \cite{Douglas:2010rc} and \cite{Koch:2010cy,Jevicki:2011ss,Jevicki:2011aa,deMelloKoch:2012vc} (see also \cite{Das:2003vw} for relevant earlier work). Recently, a dS/CFT version of the duality was also proposed \cite{Anninos:2011ui}, and further studied in \cite{Ng:2012xp, Das:2012dt}. There has also been exciting development in the $AdS_3/CFT_2$ version of higher spin holography (see \cite{Gaberdiel:2012uj} and references therein), as well as in higher dimensions \cite{Didenko:2012vh}. In this paper we will not attempt a comprehensive review of all of the dualities, their evidences and implications, but rather focus on a self-contained review of the calculation of correlation functions in Vasiliev theory and comparison with the dual vector models. We will also present a few new computations, including partial results on the parity odd terms in the three-point functions of currents in parity violating Vasiliev theory - a key recipe in identifying the holographic dual of Chern-Simons vector models.

In the next section we review the frame-like formalism of Vasiliev's higher spin gauge theory in $AdS_4$. We will mostly discuss the bosonic theory, and how the spectrum of higher spin fields arise from the linearized equations around the $AdS_4$ vacuum. We will then briefly describe the non-abelian and supersymmetric generalizations. In section 3, we review the conjectured dualities between the parity invariant Vasiliev theory and bosonic/fermionic free/critical large $N$ vector models, and their parity violating generalizations. Section 4 formulates the perturbation theory of Vasiliev's system and describes the general strategy in computing boundary correlation functions. The explicit computation of three-point functions is presented in section 5. We will discuss two approaches: by solving Vasiliev's equations in the physical spacetime, and by gauging away the spacetime dependence and work in the ``$W=0$ gauge" (gauge function method). The former approach is somewhat messy and thus far only partial results on the correlation functions are extracted, and agreement is found with the conjectured dual theories, including parity {\it odd} contributions to three-point functions involving one scalar operator. The latter approach is in principle simpler, but appears to be singular and a contour prescription is employed to regularize the calculation, which produces fully the three point functions of currents of all spins that agree with the dual CFT in the parity invariant case. We summarize the results in section 6, and discuss open questions and puzzles in section 7.

\section{Vasiliev's higher spin gauge theory}

Vasiliev's system is a set of classical nonlinear gauge invariant equations for an infinite tower of higher spin gauge fields in $AdS_4$. The equations are most conveniently written in the frame-like formalism, where the higher spin fields are packaged into connection one-forms that take value in the higher spin algebra (in the present case, the universal enveloping algebra of the $AdS_4$ isometry algebra $so(3,2)$), along with infinitely many auxiliary fields. Let us note that the equations can in principle be expressed in terms of the metric-like symmetric tensor fields as well, such that the linearized equations take the standard Fronsdal form, though in practice this procedure can only be implemented order by order in perturbation theory and is very cumbersome. We will first describe Vasiliev's master fields and equations in the frame-like formalism, and then show how the propagating degrees of freedom and the equations for metric-like fields can in principle be recovered. For comprehensive reviews on Vasiliev's higher spin gauge theory see e.g. \cite{Vasiliev:1995dn, Vasiliev:1999ba, Bekaert:2005vh, Iazeolla:2008bp}.

\subsection{Coordinates and star product}

We will denote by $x^\mu$ coordinates on the four-dimensional spacetime manifold. In addition, one introduces an internal ``twistor space", parameterized by two sets of commuting spinor variables, $(Y,Z)=(y^\A,\bar y^\da, z^\A, \bar z^\da)$, where $\A=1,2$ and $\da = \dot 1, \dot 2$. In Lorentzian signature, $(y,z)$ and $(\bar y, \bar z)$ are complex conjugates of one another. In Euclidean signature, writing the local rotation group as $SU(2)_L\times SU(2)_R$, $y, z$ transform as spinors of $SU(2)_L$, and $\bar y, \bar z$ spinors of $SU(2)_R$. Vasiliev's equations are formulated in terms of master fields that depend on both the spacetime coordinates $x^\mu$ and the internal twistor variables $(Y,Z)$, and with a non-commutative star product on the internal twistor space. 

Given two functions of the twistor variables, $f(Y,Z)$ and $g(Y,Z)$, their star product is defined as
\ie
&f(Y,Z)* g(Y,Z) = f(Y,Z) \exp\left[ \epsilon^{\A\B} \left(\overleftarrow\partial_{y^\A} + \overleftarrow\partial_{z^\A}\right) \left(\overrightarrow\partial_{y^\B} - \overrightarrow\partial_{z^\B}\right) \right.
\\
&\left.~~~~~~+ \epsilon^{\da\db} \left(\overleftarrow\partial_{\bar y^\da} + \overleftarrow\partial_{\bar z^\da}\right) \left(\overrightarrow\partial_{\bar y^\db} - \overrightarrow\partial_{\bar z^\db}\right) \right] g(Y,Z).
\fe
Note that the star product between the holomorphic varibles $(y,z)$ and the anti-holomorphic variables $(\bar y,\bar z)$ is simply the ordinary product. We have the star commutators
\ie{}
[y^\A , y^\B]_* = 2 \epsilon^{\A\B},~~~ [z^\A , z^\B]_* = - 2\epsilon^{\A\B},~~~ [y^\A , z^\B]_* = 0.
\fe
While the star product on functions of $(y,\bar y)$ only (or $(z,\bar z)$ only) takes the form of a Moyal product, the star product of functions that depend on both $Y$ and $Z$ is not quite the same as a Moyal product. In particular, the star contraction between $y$ and $z$ is nonzero, despite that they $*$-commute.\footnote{Formally, nonetheless, Vasiliev's star product appears isomorphic to a Moyal product, via the map
\ie
f(Y,Z) \mapsto {\cal O}_f(Y,Z) = \exp\left( \epsilon^{\A\B}\partial_{y^\A}\partial_{z^\B} + \epsilon^{\da\db}\partial_{\bar y^\da}\partial_{\bar z^\db} \right) f(y,z).
\fe
We then have
\ie
{\cal O}_{f*g} = {\cal O}_f \star^M {\cal O}_g,
\fe
where $\star^M$ stands for the Moyal product, defined by
\ie
{\cal O}_1 \star^M {\cal O}_2 = {\cal O}_1 \exp\left[ \epsilon^{\A\B}\left( \overleftarrow\partial_{y^\A} \overrightarrow\partial_{y^\B}- \overleftarrow\partial_{z^\A} \overrightarrow\partial_{z^\B}\right) + \epsilon^{\da\db} \left(\overleftarrow\partial_{\bar y^\da} \overrightarrow\partial_{\bar y^\db} - \overleftarrow\partial_{\bar z^\da} \overrightarrow\partial_{\bar z^\db}\right) \right] {\cal O}_2 .
\fe
Naively, Vasiliev's equations would seem to simplify when written in terms of the Moyal $\star$-algebra. Unfortunately, the propagators for the master fields appear to be singular when mapped to the Moyal $\star$-algebra, and cannot be used to compute correlators directly.}

Note that bilinears of $(y,\bar y)$ generate the Lie algebra $so(3,2)$ under star commutators, and even functions of $(y,\bar y)$ generate the higher spin algebra in three dimensions. We will see that the variables $(z,\bar z)$ are purely auxiliary: they are useful in writing the equations of motion in terms of the master fields, but all physical degrees of freedom will be contained in the master fields restricted to $z_\A = \bar z_\da =0$.

It is often convenient to use the following integral representation of the star product,
\ie
&f(y,\bar y, z,\bar z) * g(y,\bar y,z,\bar z) \\
&= \int d^2u d^2\bar u d^2v d^2\bar v \, e^{ uv+\bar u\bar v }
f(y+u,\bar y + \bar u,z+u,\bar z+\bar u) g(y+v,\bar y+\bar v,z-v,\bar z-\bar v)
\fe
where the integration measure is normalized so that $1*f=f*1=f$, and an appropriate contour prescription is used to ensure the convergence of the integral. Namely, we assume that $(u^\A, v^\A)$ are integrated along the contour $e^{\pi i/4}\mathbb{R}$ in the complex plane, and $(\bar u^\da, \bar v^\da)$ along the contour $e^{-\pi i/4}\mathbb{R}$. 

It is useful to introduce the functions
\ie
K(t)=e^{t z^\alpha y_\alpha} ,~~~~ \overline K(t) = e^{t \bar z^\da \bar y_\da}.
\fe
The special cases $K(1)\equiv K$ and $\overline{K}(1)\equiv\overline{K}$ are called Kleinians. They have the following property when starred with a general function,
\ie
f(y,z) * K=f(-z, -y) K , ~~~~ K * f(y,z)=K f(z,y).
\fe
and obey
\ie
K * K= \overline K * \overline K = 1,~~~K = \delta^2(y)*\delta^2(z),~~~\overline K = \delta^2(\bar y)*\delta^2(\bar z).
\fe

\subsection{Master fields and equations of motion}

Vasiliev's master fields are a 1-form $W=W_\mu dx^\mu$ in $x$-space, a 1-form $S=S_\A dz^\A + S_\da d\bar z^\da$ in $Z$-space, and a scalar $B$, all of which depend on spacetime coordinates $x^\mu$ as well as the internal twistor coordinates $(Y,Z)$. One may also combine $W$ and $S$ into a single 1-form ${\cal A}=W_\mu dx^\mu + S_\A dz^\A + S_\da d\bar z^\da$ on $(x,Z)$-space. The master fields are further subject to the truncation condition
\ie\label{Rtrunc}
{}[R, W]_* = \{R, S\}_*=[R, B]_*=0.
\fe
where $R\equiv K \overline K$. In other words, $W_\mu$ and $B$ are even functions of $(Y,Z)$, whereas $S_\A$ and $S_\da$ are odd functions of $(Y,Z)$.

An infinitesimal gauge transformation is parameterized by a function $\epsilon(x,Y,Z)$ that obeys $[R,\epsilon]_*=0$. The gauge variations of ${\cal A}$ and $B$ are
\ie
\delta {\cal A} = d\epsilon + [{\cal A},\epsilon]_*,~~~ \delta B = - \epsilon* B + B*\pi(\epsilon).
\fe
where $d=d_x + d_Z$, $d_x$ and $d_Z$ being the exterior derivatives in $x^\mu$ and $(z^\A, \bar z^\da)$ respectively, and $\pi$ is generally defined as the operation that flips the signs of $(y,z,dz)$ while preserving the signs of $(\bar y,\bar z,d\bar z)$. Since $\epsilon$ does not involve differentials in $(z,\bar z)$, the action of $\pi$ is equivalent to conjugation by either $K$ or $\overline K$, namely $\pi(\epsilon) = K * \epsilon * K = \overline K* \epsilon * \overline K$. ${\cal A}$ can be regarded as a $*$-algebra valued connection 1-form, whereas the scalar master field $B$ transforms in the twisted adjoint representation of the $*$-algebra. Note that $B*K$ and $B*\overline K$ transform in the adjoint representation. 

It is useful to also define
\ie
\hat{\cal A}&= {\cal A} + {1\over 2} z_\A dz^\A + {1\over 2} \bar z_\da d\bar z^\da\\
&=W_\mu dx^\mu 
+({1\over 2} z_\alpha +S_\alpha) dz^\alpha +({1\over 2} {\bar z}_{{\dot \alpha}}+S_{{\dot \alpha}})
d{\bar z}^{{\dot \alpha}},
\fe
so that the exterior derivative in $Z$ is now absorbed into the commutator with $\hat {\cal A}$.

The general gauge invariant equations of motion of Vasiliev's system takes the form
\ie\label{veqns}
&d_x \hat {\cal A} + \hat {\cal A}*\hat {\cal A} = f_*(B*K) dz^2 + \overline f_*(B*\overline K) d\bar z^2,
\\
&d_x B + \hat{\cal A}*B-B*\pi(\hat{\cal A})=0 .
\fe
Here $f(X)$ is an analytic function of $X$, and $\overline f$ its complex conjugate. $f_*(X)$ is the corresponding $*$-function, that is, replacing all products of $X$ in the Taylor series of $f(X)$ by $*$-products. Note that with a generic non-degenerate function $f(X)$, the second equation for $B$ is equivalent to the Bianchi identity that follows from the first equation. The consistency of these equations depends crucially on the fact that we have only two $z^\A$'s and two $\bar z^\da$'s (so that there is no holomorphic 3-form in $z$), and the truncation condition (\ref{Rtrunc}). The function $f(X)$ reflects some freedom in the interactions allowed by higher spin gauge symmetry, and will be discussed in the next subsection.

One may impose a further truncation on the master fields, by demanding
\ie\label{mintrunc}
& W(x,iy,i\bar y,-iz,-i\bar z) = -W(x,y,\bar y,z,\bar z), \\
& S (x,iy,i\bar y,-iz,-i\bar z,-idz,-id\bar z) = -S (x,y,\bar y,z,\bar z,dz,d\bar z),\\ 
& B(x,iy,-i\bar y,-iz,i\bar z) = B(x,y,\bar y,z,\bar z).
\fe
The resulting theory is known as the ``minimal bosonic theory".
To see that (\ref{mintrunc}) is a consistent truncation of the equations of motion, consider the involution $\iota_\pm$ on the $*$-algebra defined by sending $(y,\bar y,z,\bar z)\mapsto (iy,\pm i\bar y,-iz,\mp i\bar z)$ and reversing the order of the $*$ products. It preserves $*$-algebra in the sense that
\ie
\iota_\pm (f* g) = \iota_\pm (g)* \iota_\pm (f),
\fe
for any functions $f$ and $g$. There is also 
\ie
\iota_+ (B*K) = \iota_-(B)*K,~~~\iota_+(B*\overline K) = \iota_-(B)*\overline K.
\fe
The minimal truncation can be expressed as
\ie
\iota_+(W) = -W, ~~~ \iota_+(S) = - S,~~~ \iota_-(B) = B.
\fe
We see that it is indeed consistent with the equations of motion (\ref{veqns}).

While the original Vasiliev system, as we will see, describes interacting higher spin gauge fields of spins $s=0,1,2,3,\cdots$, the truncation to the minimal bosonic theory retains only the fields of even spins. We will mostly work with the non-minimal bosonic theory that contains all non-negative integer spins; analogous results for the minimal theory can then be extracted easily.

\subsection{Parity}

While the choice of function $f(X)$ gives an infinite parameter family of Vasiliev theories in $AdS_4$, not all $f(X)$ define physically distinct theories. The following field redefinitions
\ie
& B\to g_*(B*K)*K,
\\
& \widehat S_z \equiv ({1\over 2} z_\A + S_\A) dz^\A \to \widehat S_z * h_*(B*K),
\\
& \widehat S_{\bar z} \equiv ({1\over 2} \bar z_\da + S_\da) dz^\A \to \widehat S_{\bar z} * \overline h_*(-B*\overline K),
\fe
where $g(X)$ is any {\sl odd real} function $g(X)$, and $\bar h$ the complex conjugate of $h$, $h(X)$ being an invertible complex function, preserve the form of the gauge transformations and the equations of motion, and are consistent with the reality condition on the fields. One can use these field redefinitions to put $f(X)$ in the form
\ie\label{fxsol}
f(X) = \frac{1}{4}+X \exp(i\theta(X))
\fe
where 
\begin{equation}
\theta(X)=\sum_{n=0}^\infty \theta_{2n} X^{2n}
\end{equation}
is a real even analytic function.  The function (\ref{fxsol}), or the phase $\theta(X)$, characterizes the general parity-violating Vasiliev theory.

Parity symmetry acts on $(Y,Z)$ by $y_\A\leftrightarrow \bar y_\da$, $z_\A\leftrightarrow \bar z_\da$, and so exchanges the two terms $f(B*K)dz^2$ and $\overline f(B*\bar K)d\bar z^2$ in the equation of motion. If we imposes parity symmetry, we may assign $B$ to be either parity even or odd, and demand respectively $f(X)=f^*(X)$ or $f(X)=f^*(-X)$. In the two cases, we must have either
\ie
f_A(X)=\frac{1}{4}+X,~~{\rm or}~~f_B(X)=\frac{1}{4}+iX,
\fe
i.e. the phase $\theta(X)$ is either identically zero or equal to $\pi/2$. They define the A-type and B-type theories, respectively. If $\theta(X)$ is not identically $0$ or $\pi/2$, parity symmetry is explicitly broken.

\subsection{$AdS_4$ vacuum}

So far we have formulated Vasiliev's system in a background independent manner. It is not at all obvious that the equations (\ref{veqns}) describe higher spin gauge fields in $AdS_4$. To formulate perturbation theory, one must expand the fields around a given background that solves the equation of motion. The maximally symmetric, vacuum, solution describing $AdS_4$ spacetime takes the form 
\ie
&{\cal A} = W_0(x|Y) = e_0(x|Y) + \omega_0(x|Y) \\
&~~= (e_0)_{\A\db} y^\A\bar y^\db + (\omega_0)_{\A\B}y^\A y^\B + (\omega_0)_{\da\db} \bar y^\da \bar y^\db,~~~B=0.
\label{AdS4-master}
\fe
Here $e_0$ and $\omega_0$ are the vierbein and spin connection 1-forms on $AdS_4$. They are related to the standard vierbein and spin connection $e^a,\omega^{ab}$ in $SO(4)$ vector notations by 
\begin{equation}\label{trans}
(e_0)_{\alpha {\dot \beta}}= \frac{1}{4} e^a \sigma^a_{\alpha {\dot \beta}},~~~~
(\omega_0)_{\alpha \beta}=\frac{1}{16} \omega^{ab} \sigma^{ab}_{\alpha \beta},~~~~
(\omega_0)_{\da \db}=-\frac{1}{16} \omega^{ab} {\bar \sigma}^{ab}_{\da \db}.
\end{equation}
Indeed, with $S$ and $B$ set to zero, the only nontrivial component of Vasiliev's equation is
\begin{equation}
d_x W_0+W_0*W_0=0\,.
\end{equation} 
Collecting the independent terms in $(y,\bar y)$, one finds the equations
\begin{equation}
\begin{split}
y^{\alpha} \bar{y}^{\dot{\alpha}} &:~~~~ d_x e_{\alpha\dot{\beta}} + 4 \omega_{\alpha}^{~\beta}\wedge e_{\beta\dot{\beta}} 
          - 4 e_{\alpha\dot{\gamma}} \wedge \omega^{\dot{\gamma}}_{~\dot{\beta}} = 0, \\
y^{\alpha} y^{\beta} &:~~~~ d_x \omega_{\alpha}^{~\beta} - 4 \omega_{\alpha}^{~\gamma}\wedge w_{\gamma}^{~\beta} 
                  - e_{\alpha\dot{\alpha}}\wedge e_{\beta\dot{\beta}}\epsilon^{\dot{\alpha}\dot{\beta}} = 0, \\ 
y^{\dot{\alpha}}y^{\dot{\beta}} &:~~~~ d_x \omega^{\dot{\alpha}}_{~\dot{\beta}} 
                 + 4 \omega^{\dot{\alpha}}_{~\dot{\gamma}}\wedge \omega^{\dot{\gamma}}_{~\dot{\beta}} 
               - e_{\alpha\dot{\alpha}}\wedge e_{\beta\dot{\beta}}\epsilon^{\alpha\beta} = 0, \\ 
\end{split}
\label{3eqn}
\end{equation} 
or in vector notations,
\begin{equation}
\begin{split}
&d_x e_a + \omega_{ab}\wedge e_b = 0 \\
&d_x \omega_{ab} + \omega_{ac}\wedge \omega_{cb} 
              + 6 e_a \wedge e_b = 0.\\
\end{split}
\label{TR}	
\end{equation}
The first equation says the torsion vanishes, while the second equation relates the Riemann curvature 2-form to the vierbein and implies that the solution is the maximally symmetric space $AdS_4$. 

In Poincar\'e coordinates, with the metric given in Euclidean signature by
\ie
ds^2 = {d\vec x^2+dz^2\over z^2},
\fe
we can write $\omega_0$ and $e_0$ explicitly as
\ie
&\omega_0(x|Y) = -{1\over  8} {dx^i\over z}\left( y\sigma^{iz}y + \bar y\sigma^{iz}\bar y \right),\\
& e_0(x|Y) = -{1\over 4} {dx_\mu\over z} y\sigma^\mu \bar y.
\fe
Our convention for spinor contraction is such that the upper left spinor index is always contracted with a lower right index. The indices are raised with $\epsilon^{\A\B}$ and lowered with $\epsilon_{\A\B}$. The $\sigma_\mu$ matrices are assigned the index structure $(\sigma_\mu)_{\A\db}$, and are related to the $SO(4)$ Gamma matrices by
$$
\gamma_\mu = \begin{pmatrix} 0 & \sigma_\mu \\ \overline\sigma_\mu & 0 \end{pmatrix},
$$
where $(\overline\sigma_\mu)^{\da\B} \equiv \epsilon^{\da\dc} \epsilon^{\B\D} (\sigma_\mu)_{\D\dc}$. In contracting with chiral or anti-chiral spinors, we will not distinguish between $\overline\sigma$ and $\sigma$, with the understanding that the indices are raised or lowered by $\epsilon$ symbol as necessary. For instance, we have $y\sigma^\mu \bar y = y^\A (\sigma^\mu)_\A{}^\db \bar y_\db = \bar y^\db (\sigma^\mu)^\A{}_{\db} y_\A = \bar y \sigma^\mu y$, etc.

\subsection{Linearized equations}
\label{LinEq-sec}
The perturbation theory can be formulated by expanding the master fields around the $AdS_4$ vacuum,
\ie
W = W_0(x|Y) + \widehat W(x|Y,Z),~~~ S = S(x|Y,Z), ~~~ B = B(x|Y,Z),
\fe
and solving Vasiliev's equations order by order. At the linearized level, the equations are
\ie\label{lineareqn}
& D_0\widehat W = 0,
\\
& d_Z \widehat W + D_0 S = 0,
\\
& d_Z S = e^{i\theta_0} B*K dz^2 + e^{-i\theta_0} B*\overline K d\bar z^2,
\\
& \widetilde D_0 B = 0,
\\
& d_Z B = 0,
\fe
where $D_0$ and $\widetilde D_0$ are the covariant and twisted covariant differential with respect to $W_0$, namely
\ie
D_0 = d_x + [W_0, \cdot\,]_*,~~~~ \widetilde D_0 = d_x + [\omega_0, \cdot\,]_* + \{e_0, \cdot\,\}_*.
\fe
One finds that the Vasiliev's system describes the free propagation of a tower of massless higher spin fields, one for each integer spin,\footnote{Or one for each even spin in the minimally truncated theory.} plus a scalar with mass squared $m^2=-2$ in $AdS$ units. We will sketch the derivation of this below. As already remarked, the $Z$-twistor variable is entirely auxiliary, and the physical degrees of freedom are already contained in the master fields $\widehat W$ and $B$ restricted to $Z\equiv (z_\A,\bar z_\da)=0$. The master field $S$ is also purely auxiliary, as one may impose the gauge condition \cite{Vasiliev:1995dn, Vasiliev:1999ba}
\begin{equation}
S_{\alpha}|_{Z=0}=S_{\dot\alpha}|_{Z=0}=0\,,
\end{equation}
which can always be achieved with a $Z$-dependent gauge transformation $\epsilon(x|Y,Z)$. This gauge condition will be insisted upon in doing perturbation theory, in order to identify the higher spin fields with components of 
$$
\Omega(x|Y)\equiv \widehat W|_{Z=0}~~~{\rm  and}~~~ C(x|Y)\equiv B|_{Z=0}.
$$
In fact, the spin-$s$ degrees of freedom are contained in 
\ie
\Omega_\mu(x|Y)|_{y^{s-1+m}\bar y^{s-1-m}},~~~~C(x|Y)|_{y^{2s+n}\bar y^n},~~~~C(x|Y)|_{y^n\bar y^{2s+n}},
\fe
for $-(s-1)\leq m\leq (s-1)$ and $n\geq 0$.
In particular, $$\Omega(x|Y)|_{y^{s-1}\bar y^{s-1}} = \Omega_{\mu|\A_1\cdots \A_{s-1}\db_1\cdots\db_{s-1}}(x)\, y^{\A_1}\cdots y^{\A_{s-1}} \bar y^{\db_1}\cdots \bar y^{\db_{s-1}}dx^{\mu}$$ is the spin-$s$ field in frame-like form. After a partial gauge fixing, it is related to the metric-like rank-$s$ symmetric traceless tensor gauge field by\footnote{To write the linearly gauge invariant Fronsdal equation with rank-$(s-1)$ symmetric traceless tensor gauge parameter, one shall allow the spin-$s$ gauge field to be double-traceless rather than traceless. The trace part can be gauged away, nonetheless.}
\ie
\Omega_{\mu|\A_1\cdots\A_{s-1}\db_1\cdots\db_{s-1}} = \sigma^{\nu_1}_{\A_1\db_1}\cdots\sigma^{\nu_{s-1}}_{\A_{s-1}\db_{s-1}} \Phi_{\mu \nu_1\cdots\nu_{s-1}}.
\fe
The other components $\Omega(x|Y)|_{y^{s-1+m}\bar y^{s-1-m}}$ contain the higher spin version of the spin connection, and are related to the frame field by the equations of motion. The components of the scalar master field that depend on only $y$ or $\bar y$, namely $C|_{y^{2s}}, C|_{\bar y^{2s}}$, contain the self-dual and anti-self-dual parts of the higher spin generalization of the Weyl curvature tensor (for $s=2$ they are the standard self-dual and anti-selfdual parts of the gravity Weyl tensor). Finally, the bulk scalar field is given by the bottom component of $C(x|Y)$, namely $C(x|y=\bar y=0)$. All the higher components $C(x|Y)|_{y^{2s+n}\bar y^n}, C(x|Y)|_{y^n\bar y^{2s+n}}$ with $n>0$ are related to the lowest component $n=0$ by the equations of motion. 

\subsubsection{The scalar field}

Now let us derive the linearized equation for the scalar field from the equation of the $B$ master field explicitly. The last line of the linearized equation (\ref{lineareqn}) implies that $B^{(1)}(x|Y,Z) = C^{(1)}(x|Y)$ is independent of the twistor variable $Z$. Now let us analyze the fourth equation in (\ref{lineareqn}) for $B$, or $C(x|Y)$, which may be written as
\ie\label{ceqn}
&\widetilde D_0 C(x|Y) = \nabla_\mu^L C(x|Y) + \{ e_\mu^{\A\db} y_\A \bar y_\db, C(x|Y) \}_* 
\\
&= \nabla_\mu^L C(x|Y) +2 e_\mu^{\A\db} y_\A \bar y_\db C(x|Y) + 2e_\mu^{\A\db} \partial_{y^\A} \partial_{\bar y^\db} C(x|Y)  = 0,
\fe
where $\nabla_\mu^L = \partial_\mu + [(\omega_0)_\mu, \cdot\,]_*$.
Denote by $C^{(n,m)}$ the terms in $C(x|Y)$ of degree $n$ in $y$ and $m$ in $\bar y$. Note that $\nabla_\mu^L$ does not change the degree in $y$ nor $\bar y$. We see that (\ref{ceqn}) only relates $C^{(n,m)}$'s with the same $n-m$. As already mentioned, the scalar field is contained in $C^{(n,n)}$, $n\geq 0$. To obtain the second order linearized equation for $C^{(0,0)}$, it will suffice to consider the $(0,0)$ and $(1,1)$ components of (\ref{ceqn}),
\ie\label{ceqnb}
& \partial_\mu C^{(0,0)}(x) + 2e_\mu^{\A\db} \partial_{y^\A} \partial_{\bar y^\db} C^{(1,1)}(x|Y)  = 0,
\\
& \nabla_\mu^L C^{(1,1)}(x|Y) +2 e_\mu^{\A\db} y_\A \bar y_\db C^{(0,0)}(x) + 2e_\mu^{\A\db} \partial_{y^\A} \partial_{\bar y^\db} C^{(2,2)}(x|Y)  = 0.
\fe
The higher order components of the equation are then guaranteed to be satisfied by the compatibility of the equations. Expanding
\ie
C^{(1,1)}(x|Y) = C^{(1,1)}_{\A\db}(x) y^\A \bar y^\db,~~~~ C^{(2,2)}(x|Y) = C^{(2,2)}_{\A\B\dc\dd}(x) y^\A y^\B \bar y^\dc\bar y^\dd,
\fe
we have $C^{(1,1)}_{\A\db} = 4 e^\mu_{\A\db}\partial_\mu C^{(0,0)}$ (in our normalization convention, $e^{\A\db}_\mu e^\mu_{\C\dd} = -{1\over 8} \delta^\A_\C\delta^\db_\dd$, and $e^{\A\db}_\mu e^\nu_{\A\db} = -{1\over 8}\delta^\nu_\mu$.) Contracting the second equation of (\ref{ceqnb}) with $(e^\mu)^{\A\db}$, we have
\ie
& (e^\mu)^{\A\db} \nabla_\mu^L C^{(1,1)}(x|Y) -{1\over 4} y^\A \bar y^\db C^{(0,0)}(x) - y_\C \bar y_\dd C^{(2,2)}{}^{\A\C\db\dd}(x|Y)  = 0.
\fe
Now acting on this equation with $\partial_{y^\A} \partial_{\bar y^\db}$, $C^{(2,2)}$ drops out, and we obtain
\ie
& (e^\mu)^{\A\db}\partial_{y^\A} \partial_{\bar y^\db} \nabla_\mu^L C^{(1,1)}(x|Y) - C^{(0,0)}(x|Y) = 0,
\fe
or
\ie
&4 (e^\mu)^{\A\db}\partial_{y^\A} \partial_{\bar y^\db} \nabla_\mu^L \left[ e^\nu_{\C\dd} y^\C \bar y^\dd \partial_\nu C^{(0,0)}(x) \right] - C^{(0,0)}(x) 
\\
&= - {1\over 2} \partial_\mu\partial^\mu C^{(0,0)}(x) + 8 (e^\mu)^{\A\db} \left[ (\omega_0)_\mu{}_{\A\C'} \epsilon^{\C'\C} e^\nu_{\C\db} + (\omega_0)_\mu{}_{\dd'\db} \epsilon^{\dd'\dd} e^\nu_{\A\dd} \right]  \partial_\nu C^{(0,0)}(x) - C^{(0,0)}(x)  
\\
& = -{1\over 2} \left(\nabla_\mu \partial^\mu +2 \right) C^{(0,0)}(x)=0,
\fe
where $\nabla_\mu$ is the ordinary covariant derivative. This is indeed the Klein-Gordon equation for a massive scalar of mass squared $m^2=-2$ in $AdS$ units.

We could derive analogous equations for $C^{(2s,0)}$ and $C^{(0,2s)}$. Though, as remarked, these are the higher spin Weyl curvature, related to up to $s$ derivatives of the fundamental spin-$s$ symmetric traceless tensor field, which is contained in $\Omega^{(s-1,s-1)}$.

\subsubsection{The higher spin fields}

To derive the linearized equations for the fields of nonzero spins, we must examine the first three equations of (\ref{lineareqn}). It is useful to split $\widehat W$ into two parts,
\ie
\hat W(x|Y,Z) = \Omega(x|Y) + W'(x|Y,Z),
\fe
with $W'|_{Z=0}=0$. $W'$ is then entirely determined by $S$, which is further determined by $B$ when the gauge condition $S|_{Z=0}=0$ is imposed. On the other hand, the first equation of (\ref{lineareqn}) relates $\Omega$ to $W'$, namely,
\ie
D_0 \Omega = -D_0 W' = - (D_0 W')|_{Z=0} = - \{W_0, W'\}_*|_{Z=0}.
\fe
Let us now analyze these equations in detail. Begin with the linearized field $B(x|Y,Z) = C(x|Y)$. Using the linearized equation
$d_Z S = e^{i\theta_0} B*K dz^2+e^{-i\theta_0} C*\overline K d\bar z^2$, we can solve for $S$ by integrating in $(z,\bar z)$,
\ie\label{sbrel}
S_\A &= - e^{i\theta_0} z_\A d \int_0^1 dt \,t (C*K)|_{z_\A\to t z_\A}  \\
&= - e^{i\theta_0} z_\A \int_0^1 dt \,t\, C(x|-t z,\bar y) K(t) ,
\fe
and a similar complex conjugated expression for $S_\da$.
Recall that we have defined $K(t) = e^{t z^\A y_\A}$. Next, using $d_Z \widehat W = -D_0 S$,
we can solve for $W'$ by integrating again in $(z,\bar z)$,
\ie\label{wpr}
W' &=  z^\A \int_0^1 dt\, (D_0S_\A)|_{z\to tz} +c.c. 
\\
& = - e^{i\theta_0} z^\A \int_0^1 dt \int_0^1 dt'\,t'\, \left[ W_0, z_\A C(x|-t'z,\bar y) K(t') \right]_* |_{z\to tz} +c.c.
\\
& = - e^{i\theta_0} z^\A \int_0^1 dt \int_0^1 dt'\,t'\, \left[ (e_0)_{\A\db} \bar y^\db + (\omega_0)_{\A\B}y^\B, C(x|-t'z,\bar y) K(t') \right]_* |_{z\to tz} +c.c.
\\
& = 2e^{i\theta_0} z^\A \int_0^1 dt\,(1-t) \left[ (e_0)_\A{}^\db \partial_{\bar y^\db} + (\omega_0)_{\A\B} t z^\B \right] C(x|-tz,\bar y) K(t) +c.c.
\fe
Now, we can relate $\Omega$ to $C$, by
\ie\label{ewaa}
& D_0 \Omega = \left.- \{ W_0, W'\}_*\right|_{Z=0} 
\\
&=\left. -4 e^{i\theta_0}  \int_0^1 dt\,(1-t) \epsilon^{\C\A}\left[ (e_0)_\A{}^\db \partial_{\bar y^\db} + (\omega_0)_{\A\B} t z^\B\right]\wedge \left[ (e_0)_\C{}^\dd \partial_{\bar y^\dd} + (\omega_0)_{\C\D} t z^\D \right] C(x|-tz,\bar y) K(t) \right|_{Z=0} +c.c.
\\
&= 2 e^{i\theta_0} \epsilon^{\A\C} (e_0)_\A{}^\db\wedge (e_0)_\C{}^\dd \partial_{\bar y^\db} \partial_{\bar y^\dd}  C(x| 0,\bar y) +2 e^{-i\theta_0} \epsilon^{\da\dc} (e_0)^\B{}_\da \wedge (e_0)^\D{}_\dc \partial_{y^\B} \partial_{y^\D}  C(x| y,0).
\fe
The linearized equation of motion for all fields of nonzero spin follows from (\ref{ewaa}). It also follows from (\ref{ewaa}) that $C(x|y,0)$ and $C(x|0,\bar y)$ are the self-dual and anti-self-dual parts of the higher spin Weyl curvature. We only sketch the derivation here. First, we can write $D_0 = d^L + [e_0, \cdot\,]_*$, where $d^L = d_x + [\omega_0, \cdot\,]_*$. Note that $d^L$ does not change the degree in $y$ nor $\bar y$, whereas the commutator with $e_0$ shifts the degree in $y$ or $\bar y$ by 1, while maintaining the total degree. Therefore, (\ref{ewaa}) relates components of $\Omega$ of the form $\Omega^{(s-1-k,s-1+k)}$, $-(s-1)\leq k\leq (s-1)$. It also relates $\Omega^{(2s-2,0)}$ to $C^{(2s,0)}$, and $\Omega^{(0,2s-2)}$ to $C^{(0,2s)}$.

For instance, in the spin $s=1$ case, we see that $C_{\A\B}^{(2,0)}$ is proportional to $e^{i\theta_0} F_{\A\B}$, and $C^{(0,2)}_{\da\db}$ is proportional to $e^{-i\theta_0} F_{\da\db}$, where $F_{\A\B}$ and $F_{\da\db}$ are the self-dual and anti-self-dual parts of the gauge field strength in spinorial notation. The linearized equation for the spin-1 gauge field then follows from the equation for $C^{(2,0)}$, namely
\ie
& \nabla_\mu^L C^{(2,0)}(x|Y) + 2 e_\mu^{\A\db} \partial_{y^\A} \partial_{\bar y^\db} C^{(3,1)}(x|Y)=0
\\
& \Rightarrow ~(e^\mu)^\A{}_\db \partial_{y^\A} \nabla_\mu^L C^{(2,0)}(x|Y) =0.
\fe

Note that the linearized equations for the higher spin fields are still second order. In fact, for $s>1$, the equation of motion for the spin-$s$ field already follows from
\ie\label{tempom}
d^L \Omega^{(s-1-k,s-1+k)} + 2e_0^{\A\db}\wedge \left[ y_\A\partial_{\bar y^\db} \Omega^{(s-2-k,s+k)} + \partial_{y^\A} \bar y_\db \Omega^{(s-k,s-2+k)} \right] = 0,
\fe
with $k=0,\pm 1$ only. Let us define a few shorthand notations. Denote $\Omega^{(s-1+k,s-1-k)}_\mu$ by $\Omega^k_\mu$. Next, define
\ie
& \Omega_{++}^n = (e^\mu)^{\A\db} y_\A \bar y_\db \Omega^n_\mu,
\\
& \Omega_{+-}^n = (e^\mu)^{\A\db} y_\A \partial_{\bar y^\db} \Omega^n_\mu,
\\
& \Omega_{-+}^n = (e^\mu)^{\A\db} \partial_{y^\A} \bar y_\db \Omega^n_\mu,
\\
& \Omega_{--}^n = (e^\mu)^{\A\db} \partial_{y^\A} \partial_{\bar y^\db} \Omega^n_\mu,
\label{Omega-shorthand}
\fe
so that $\Omega_\mu^n$ can be decomposed as
\ie
\Omega_\mu^n = {-8\over s^2-n^2} e_\mu^{\A\db} \left( \partial_{y^\A} \partial_{\bar y^\db} \Omega_{++}^n - \partial_{y^\A} \bar y_\db \Omega_{+-}^n - y_\A \partial_{\bar y^\db} \Omega_{-+}^n + y_\A \bar y_\db \Omega_{--}^n\right).
\fe
From (\ref{tempom}) one can then extract an expression for $\Omega^n_{\pm\pm}$ in terms of $d_L \Omega^{n\pm 1}$, of the following form:
\ie\label{omegas}
& \Omega^n_{++} = {4\over s+n-1} y^\A y^\B (\nabla^L)_\A{}^\dc \Omega^{n-1}_{\B\dc},
\\
& \Omega^n_{--} = -{4\over s-n+1} (\partial_{\bar y})^\da (\partial_{\bar y})^\db (\nabla^L)^{\C}{}_{\da} \Omega^{n-1}_{\C\db},
\\
& \Omega^n_{-+} = -{2\over s} \left[ (s-n) y^\A (\partial_y)^\B (\nabla^L)_{\A}{}^\dc \Omega^{n-1}_{\B\dc}
+ (s+n) \bar y^\da (\partial_{\bar y})^\db (\nabla^L)^\C{}_\da \Omega^{n-1}_{\C\db} \right],
\\
& \Omega^n_{+-} = -{2\over s} \left[ (s-n) y^\A (\partial_y)^\B (\nabla^L)_{\A}{}^\dc \Omega^{n+1}_{\B\dc}
+ (s+n) \bar y^\da (\partial_{\bar y})^\db (\nabla^L)^\C{}_\da \Omega^{n+1}_{\C\db} \right],
\fe
where we have defined $\nabla^L_{\A\db} \equiv e^\mu_{\A\db}\nabla^L_\mu$.

A $Z$-independent gauge transformation takes the form
\ie
\delta \Omega_\mu(x|Y) = \nabla_\mu^L \epsilon(x|Y) + 2 e_\mu^{\A\db} \left( y_\A \partial_{\bar y^\db} + \bar y^\db \partial_{y^\A} \right) \epsilon(x|Y).
\fe
Note that by choosing $\epsilon(x|Y)$ we can gauge away either $\Omega_{+-}^{(n-1)}$ or $\Omega_{-+}^{(n+1)}$ completely, for any $n$. It is convenient to partially fix the gauge $\Omega_{+-}^{(n)}=0$ for $n\geq 0$, and $\Omega_{-+}^{(n)}$ for $n\leq 0$. In the special case $n=0$, both $\Omega_{+-}^{(0)}$ and $\Omega_{-+}^{(0)}$ can be gauged away completely. 

By restricting (\ref{omegas}) to $n=1$, we can express $\Omega^{1}$ in terms of $\Omega^0$ via the relations
\ie
& \Omega^1_{++} = {4\over s} y^\A y^\B (\nabla^L)_\A{}^\dc \Omega^{0}_{\B\dc},
\\
& \Omega^1_{--} = -{4\over s} (\partial_{\bar y})^\da (\partial_{\bar y})^\db (\nabla^L)^{\C}{}_{\da} \Omega^{0}_{\C\db},
\\
& \Omega^1_{-+} = -{2\over s} \left[ (s-1) y^\A (\partial_y)^\B (\nabla^L)_{\A}{}^\dc \Omega^{0}_{\B\dc}
+ (s+1) \bar y^\da (\partial_{\bar y})^\db (\nabla^L)^\C{}_\da \Omega^{0}_{\C\db} \right].
\fe 
Combining this with the complex conjugate version of (\ref{omegas}), and in particular
\ie
& \Omega^0_{++} = {4\over s-1} \bar y^\da \bar y^\db (\nabla^L)^\C{}_\da \Omega^{1}_{\C\db},
\\
& \Omega^0_{--} = -{4\over s+1} (\partial_{y})^\A (\partial_{y})^\B (\nabla^L)_\A{}^{\dc} \Omega^{1}_{\B\dc},
\fe
one can derive a second order equation for $\Omega^0$ only. The resulting equation is equivalent to Fronsdal's equation for free higher spin gauge fields in $AdS_4$, with $\Omega^0_{++}$ and $\Omega^0_{--}$ identified with the traceless part and the trace part of the rank-$s$ symmetric double-traceless tensor field.

\subsection{Generalizations}

Having understood that Vasiliev's system describes a tower of interacting higher spin gauge fields in $AdS_4$, let us discuss two types of generalizations: to non-abelian higher spin gauge theories, and to higher spin theories with (extended) supersymmetry.

\subsubsection{Chan-Paton factors} \label{nonabelian}

Vasiliev's system in $AdS_4$ admits an obvious generalization to non-abelian 
higher spin fields, through the introduction of Chan-Paton factors, much like 
in open string field theory. All we need to do is simply to replace the master fields $W, S, B$ by 
$M\times M$ matrix valued fields, and replace the $*$-algebra in the gauge 
transformations and equations of motion by its tensor product with the algebra 
of $M\times M$ complex matrices. In making this generalization we 
modify neither the truncation condition \eqref{Rtrunc} nor the reality condition on fields, except that the complex conjugation is now defined with Hermitian conjugation on the $M\times M$ matrices. 

The spin-1 gauge field in the bulk is now a $U(M)$ non-abelian gauge field. All higher spin fields as well as the scalar transform in the adjoint representation of the $U(M)$ gauge group. One may be disturbed by the appearance of colored gravitons, which would seem to be dual to multiple stress-energy tensors in the boundary CFT. For the parity invariant theories with higher spin symmetry preserving boundary condition, the dual CFT is free with $U(M)$ flavor symmetry, thus all conserved currents transform in the adjoint of $U(M)$. When the boundary condition breaks the higher spin symmetry, as generally occurs in the parity violating theory, or in the case of the parity invariant theory with alternative boundary condition assigned to the bulk scalar, it can be seen, from either the bulk or the boundary theory, that the $SU(M)$ part of the spin-2 currents are no longer conserved. The colored gravitons become massive, and their longitudinal modes are supplied by two-particle (bound) states.

Our perturbative computation of boundary correlation functions will generalize straightforwardly to the non-abelian higher spin gauge theory. All that is needed is to attached appropriate group theory factors to the color-ordered correlators.

\subsubsection{Supersymmetric theories}

Vasiliev's system with extended supersymmetry \cite{Engquist:2002vr, Chang:2012kt} can be defined by simply introducing
Grassmannian auxiliary variables $\psi_i$, $i=1,\cdots,n$, that obey 
Clifford algebra $\{\psi_i, \psi_j\} = 2\delta_{ij}$, and commute with all 
the twistor variables $(Y,Z)$. By definition, the $\psi_i$'s do not 
participate in the $*$-algebra.  The master fields $W, S, B$, as 
well as the gauge transformation parameter $\epsilon$,  are now
functions of the $\psi_i$'s as well as of 
$(x^\mu, y_\A, \bar y_\da, z_\A, \bar z_\da)$. 

The truncation condition \eqref{Rtrunc} continues to take the form 
\ie\label{newtrunc}
{}[R,W]_*=\{R,S\}_*=[R,B]_*=[R,\epsilon]_*=0.
\fe
but with $R$ now defined as 
\begin{equation}\label{Rdef}
R\equiv K \overline K\Gamma,
\end{equation}
where 
\begin{equation} \label{gdef} 
\Gamma\equiv i^{n(n-1)\over 2} \psi_1\psi_2\cdots\psi_n.
\end{equation} 
Note that we still have $\Gamma^2=1$ and $R*R=1$. The truncation condition
is such that the components of the master fields that are even functions of $\psi_i$ are also even functions of 
the spinor variables $Y, Z$, whereas odd functions of $\psi_i$ are also odd functions of 
$Y, Z$. The latter give rise to fermionic fields in $AdS_4$ of half integer spins.

The reality condition on the fields is as follows. 
\ie \label{imdef}
\iota(W)^*=-W,~~~\iota(S)^* = -S,~~~\iota(B)^* = \overline K*B* \overline K\Gamma = \Gamma K*B*K.
\fe
The operation $\iota$ is defined as $\iota_+$ in section 2.2, combined with $\iota:~\psi_i\to\psi_i$ but reverses the order of the product of $\psi_i$'s as well as the $*$ product. $\psi_i$'s are real under complex conjugation. 

The supersymmetric extension of Vasiliev's equations takes the form
\ie\label{veqnss}
& d_x \hat {\cal A} + \hat {\cal A}*\hat {\cal A} = f_*(B*K) dz^2 + \overline f_*(B*\overline K \Gamma) d\bar z^2,
\\
&d_x B + \hat{\cal A}*B-B*\pi(\hat{\cal A})=0 .
\fe
Compared to the bosonic theory, the only change in the first 
Vasiliev equation is the factor of $\Gamma$ in the 
argument of ${\overline f}$; this factor is needed in order to preserve 
the reality of Vasiliev equations under the operation \eqref{imdef}.
The second Vasiliev equation takes an identical form as in the bosonic theory; 
however the operator $\pi$ is now taken to mean 
conjugation by $\Gamma \overline K$ together with $d\bar z\to -d\bar z$, or equivalently, by the truncation condition (\ref{newtrunc}) on the fields, 
conjugation by $K$ together with $dz\to -dz$.
As in the case of the bosonic theory, $f(X)$ can generically be 
put into the form  $f(X)=\frac{1}{4}+X \exp(i\theta(X))$ 
by a field redefinition.

\section{Holographic dualities}

\subsection{Parity invariant theories and free/critical vector models}

The A-type minimal bosonic Vasiliev theory in $AdS_4$, with fields of spin $s=0,2,4,\cdots$ and even parity assigned to the scalar field, is conjectured by Klebanov and Polyakov to be holographically dual to the free or critical $O(N)$ vector model \cite{Klebanov:2002ja}. The operator dual to the bulk scalar (of mass squared $m^2=-2$) has dimension
\ie
\Delta_\pm = {3\over 2}\pm \sqrt{{9\over 4}+m^2} = 2 {\rm~~or~~}1,
\fe
depending on the boundary condition. Near the boundary of $AdS_4$, the bulk scalar field $\varphi(\vec x,z)$ has the fall off behavior
\ie
\varphi \sim a z + b z^2 + {\cal O}(z^3).
\fe
There are two consistent conformally invariant boundary conditions: the Dirichlet type boundary condition setting $a=0$, or the Neumann type boundary condition setting $b=0$, in the absence of sources. With the former boundary condition, the dual operator has dimension $\Delta=2$, whereas in the latter case the dual operator has dimension $\Delta=1$. The holographic duality states that the A-type Vasiliev theory with $\Delta=1$ boundary condition is dual to the free $O(N)$ vector model, that is, the theory of $N$ massless scalar fields $\phi_i$ in three dimensions, restricted to the $O(N)$ singlet sector. As a CFT, the singlet condition means that the set of physical operators are $O(N)$ invariant. For instance, the bulk scalar field is dual to the dimension 1 operator ${\cal O}=\sum_{i=1}^N\phi_i\phi_i$, whereas the fundamental scalar $\phi_i$ of the boundary field theory is not in the operator spectrum. The restriction to the singlet sector can be consistently implemented for correlators on $\mathbb{R}^3$.

The set of higher spin gauge fields in the bulk are dual to single-trace conserved higher spin currents of the free $O(N)$ vector model. One can write a generating function for these currents,
\ie
\sum_{s=0}^\infty J^{(s)}_{\mu_1\cdots\mu_s}(x) \varepsilon^{\mu_1}\cdots\varepsilon^{\mu_s}
= \phi_i(x) f(\varepsilon_\mu, \overleftarrow\partial_\mu, \overrightarrow\partial_\mu ) \phi_i(x),
\fe
where the function $f(\vec\varepsilon, \vec u, \vec v)$ is defined as
\ie
f(\vec\varepsilon, \vec u, \vec v) = e^{(u-v)\cdot \varepsilon} \cos\left[ \sqrt{4 (u\cdot \varepsilon)(v\cdot\varepsilon)- 2(u\cdot v)\varepsilon^2}\right].
\fe
It is not hard to see that all single trace operators (i.e. operators involving contraction of only one pair of $O(N)$ indices) are linear combinations of the currents $J^{(s)}$ and their descendants. Multi-trace operators are dual to multi-particle states in the bulk.

The non-trivial content of the duality conjecture lies in the correlation functions. The boundary correlators obey large $N$ factorization. The $1/N$ expansion of boundary correlators should map to the perturbative expansion of the bulk theory. For instance, the spin-$s$ current with normalized two-point function takes the form $J^{(s)}\sim {1\over \sqrt{N}}\phi_i \partial^s \phi_i+\cdots$. The three-point function of these currents are of order $N^{-{1\over 2}}$. Therefore, $N^{-{1\over 2}}$ should be identified with the bulk cubic coupling constant. Note that Vasiliev's theory is formulated in terms of the equation of motion, and not explicitly in terms of an action. If we assume that the equations of motion can be derived from the variation of an action, the overall coupling constant drops out of the equations. In computing holographic correlators, the one and only one overall coupling constant must be put in by hand, associated with each (cubic) vertex. After doing so, we expect the correlators computed from the bulk theory using the AdS/CFT dictionary to match with that of the boundary CFT, order by order in $1/N$. This computation for three-point functions will be discussed explicitly later.

With the alternative, $\Delta=2$, boundary condition on the bulk scalar, the dual CFT is conjectured to be the critical $O(N)$ vector model. The latter can be obtained from the free $O(N)$ vector model by turning on a double trace deformation $(\phi_i\phi_i)^2$ and flow to the critical point in the IR. Alternatively, the critical $O(N)$ model can also be described as the UV fixed point of the nonlinear sigma model on $S^{N-1}$. Again, it is important to restrict to the $O(N)$ singlet sector, for large $N$ factorization to hold.

A mere change of boundary condition on the bulk scalar field may seem insignificant, but it modifies all correlation functions, including those involving higher spin currents only, at subleading order in $1/N$. In particular, the change of scalar boundary condition changes the boundary-to-bulk propagator as well as the {\it bulk-to-bulk} propagator of the scalar field. The latter modifies all bulk loop diagrams involving a scalar field propagator. It is shown in \cite{Hartman:2006dy, Giombi:2011ya} that the duality with critical $O(N)$ model in the $\Delta=2$ case follows from the duality with the free $O(N)$ model in the $\Delta=1$ case order by order in $1/N$. Note that although the change of scalar boundary condition modifies loop corrections to correlators, these modifications can be computed entirely using the data of correlators in the case of $\Delta=1$ boundary condition.

The $\Delta=2$ boundary condition also breaks all higher spin symmetries, and the breaking is controlled by $1/N$ \cite{Girardello:2002pp}. Correspondingly, the critical $O(N)$ vector model does not have exactly conserved higher spin currents. There are, nonetheless, higher spin conformal primaries, which are approximately conserved at large $N$ in a sense that will be made precise later.

The B-type minimal bosonic Vasiliev theory in $AdS_4$, on the other hand, is conjectured by Sezgin and Sundell \cite{Sezgin:2003pt} to be dual to the fermionic version of the free and critical $O(N)$ models. With $\Delta=2$ boundary on the bulk scalar, which is parity odd, the dual CFT is the theory of $N$ free massless real fermions in three dimensions, restricted to the $O(N)$ singlet sector. With $\Delta=1$ boundary condition, on the other hand, the dual theory is the Gross-Neveu model, again restricted to the $O(N)$ singlet sector.

The dualities for the minimal bosonic theories can be straightforwardly generalized to the non-minimal case, where the bulk theory includes fields of all non-negative integer spins $s=0,1,2,3,\cdots$. The conjectured dual CFTs are the $U(N)$ versions of the bosonic or fermionic vector models. A novel feature of the non-minimal theory is that in addition to two possible boundary conditions one can impose on the boundary scalar, there is now a one-parameter family of conformally invariant boundary conditions one can impose on the bulk spin-1 gauge field. The ordinary boundary condition on a spin-1 gauge field in $AdS_4$, dual to a conserved current on the boundary, is such that the ``magnetic" field $F_{ij}$, the indices $i,j$ labeling directions parallel to the boundary, vanishes at the boundary. More generally one can impose a mixed boundary condition, namely a linear combination of the ``electric" field $F_{zi}$ ($z$ being the Poincar\'e radial coordinate) and the ``magnetic" field $\epsilon_{ijk} F_{jk}$ vanishes at the boundary. With the mixed boundary condition, the dual CFT is obtained from the original one by gauging the global $U(1)$ flavor symmetry (dual to the bulk spin-1 gauge field), via turning on Chern-Simons coupling at some level $k$. The purely ``electric" boundary condition corresponds to $k=0$, that is, while one gauges the boundary flavor current, and the kinetic term for the boundary gauge field is entirely generated from integrating out the matter fields at one-loop, as in the case of three-dimensional critical QED.

So in particular, three-dimensional critical QED with $N$ bosonic or fermionic flavors, as well as the critical $\mathbb{CP}^{N-1}$ model, restricted to $U(N)$ singlet sector, are holographically dual to A-type or B-type non-minimal Vasiliev theory, with the dual ``electric" boundary condition imposed on the bulk spin-1 gauge field.

\subsection{Parity violating theories and Chern-Simons vector models}

As already discussed in the construction of Vasiliev's system, if we relax parity symmetry, then there are more general Vasiliev theories in $AdS_4$ with the same higher spin gauge field content but with interactions controlled by the function $f(X)$, or the phase $\theta(X) = \theta_0 + \theta_2 X^2 + \theta_4 X^4+\cdots$. As will become clearly in the perturbative computation of correlation functions, $\theta_n$ controls $(3+n)$-th order and higher couplings in the bulk. In particular, the cubic coupling of higher spin gauge fields is controlled by the interaction phase $\theta_0$ and the overall bulk coupling constant $g$.

If the parity violating Vasiliev theories make sense as full quantum theories of higher spin gravity in $AdS_4$, one could ask what are the dual three-dimensional CFTs which must be large $N$ vector model like theories. Indeed, in three dimensions one can construct a larger class of vector models by coupling $N$ complex (or real) massless scalars or fermions to $U(N)$ (or $O(N)$) Chern-Simons gauge fields at some level $k$. While Chern-Simons-matter theories generally give rise to CFTs with any gauge group and any matter representation \cite{Gaiotto:2007qi}, Chern-Simons vector models are special in that the operator spectrum is not renormalized at infinite $N$ and finite 't Hooft coupling $\lambda = N/k$ \cite{Giombi:2011kc, Aharony:2011jz}. Further, in Chern-Simons vector models, all single trace operators lie in the conformal families of a single tower of primary currents of spins $s=0,1,2,3,\cdots$ (or only even spins for the $O(N)$ theories), and this set of higher spin currents are approximately conserved at large $N$. This indicates that the holographic dual of large $N$ Chern-Simons vector model must be a higher spin {\it gauge} theory, in that the classical bulk equation of motion must respect higher spin gauge symmetry so that no extra longitudinal degrees of freedom are introduced. It is then natural to conjecture that the $U(N)$ (or $O(N)$) Chern-Simons vector model is holographically dual to a one-parameter family of parity violating non-minimal (or minimal) bosonic Vasiliev theories. The interaction phases $\theta_0, \theta_2, \cdots$ should be functions of the Chern-Simons 't Hooft coupling $\lambda$. A two-loop calculation of correlation functions \cite{Giombi:2011kc}, an exact calculation in the Chern-Simons-scalar vector model at large $N$ \cite{OferUpcoming}, combined with arguments of \cite{Maldacena:2012sf, Chang:2012kt} suggests that the identification between $\theta_0$ and $\lambda$ is simply
\ie
\theta_0 = {\pi\over 2}\lambda
\fe
for Chern-Simons-scalar vector model, and
\ie
\theta_0 = {\pi\over 2}(1-\lambda)
\fe
for Chern-Simons-fermion vector model.
The relation between $\theta_2, \theta_4,\cdots$ and $\lambda$ are not yet known. If the $\theta_n$'s for $n>0$ are absent, the conjectured duality would also imply a level-rank type duality between Chern-Simons-scalar and Chern-Simons-fermion vector models (``3d bosonization" of \cite{OferUpcoming}). However, such a duality seems to contradict exact results of finite temperature free energy of these theories at infinite $N$, computed by solving Schwinger-Dyson equations in lightcone gauge \cite{Giombi:2011kc, Jain:2012qi}.

The duality between parity violating Vasiliev theory and Chern-Simons vector models is further generalized to the supersymmetric case in \cite{Chang:2012kt}. In fact, Chern-Simons vector models of ${\cal N}=0,1,2,3,4$ or $6$ supersymmetries can all be thought of as bosonic $U(N)$ Chern-Simons theory coupled to massless scalars and fermions, with additional double-trace and triple-trace terms turned on, and with the possibility of further gauging a flavor group with another Chern-Simons gauge field. Correspondingly, their holographic duals differ essentially only by a change of boundary condition, on the bulk scalars, spin-${1\over 2}$ fermions, and spin-1 gauge fields.
What is remarkable is that one can precisely identify the boundary conditions in the supersymmetric parity violating Vasiliev theories that preserve various fractions of supersymmetries, and hence the holographic duals of all supersymmetric Chern-Simons vector models. Some of these, say the ${\cal N}=6$ vector model, is a limit of ABJ theory with a known string theory dual. The higher spin/vector model duality in this case then suggests a direct bulk-bulk duality between non-Abelian supersymmetric parity violating Vasiliev theory and type IIA string field theory on $AdS_4\times\mathbb{CP}^3$.

\subsection{Boundary conditions and symmetry breaking}

With generic parity breaking phase $\theta_0$, neither $\Delta=1$ nor $\Delta=2$ boundary condition on the bulk scalar can preserve the higher spin symmetries. In other words, the higher spin symmetries are always broken, as expected from the dual interacting CFT. From the bulk perspective, this can be seen from the fact that, while the global higher spin symmetries are generated by gauge parameters $\epsilon_0(x|Y)$ that preserve the $AdS_4$ vacuum solution, namely
\ie
D_0 \epsilon_0(x|Y)=0,
\fe
the corresponding gauge transformation, say on the scalar master field $B(x|Y,Z)$,
\ie
\delta_{\epsilon_0} B = - \epsilon_0 * B + B*\pi(\epsilon_0)
\fe
may not respect the boundary conditions assigned on $B$. In fact, since the higher spin gauge transformations mix fields of different spins, starting with a spin-$s$ field (say $s>1$) that obeys its boundary condition, under the would-be symmetry generated by $\epsilon_0$, there is a nonzero variation of the bulk scalar field, $\delta_{\epsilon_0} B^{(0,0)}(x|Y)$. While the boundary condition on the spin-$s$ field is unambiguously fixed, so far we have not said whether $\Delta=1$ or $\Delta=2$ boundary condition is imposed on the scalar. Generically then, either or both boundary conditions will be violated by the higher spin symmetry variation on the scalar field. This indeed occurs for generic phase $\theta_0$ with any boundary condition, and for $\theta_0=0$ with $\Delta=2$ boundary condition, or $\theta_0 = \pi/2$ with $\Delta=1$ boundary condition, leading to the breaking of global higher spin symmetry in the dual conformal field theory.

From the boundary perspective, the breaking of higher spin symmetry (or other symmetries such flavor or supersymmetry) by $AdS$ boundary condition corresponds to the statement that the divergence of the symmetry current is a multi-trace operator at large $N$. If we normalize all higher spin currents $J^{(s)}$ so that their two-point functions do not depend on $N$ (or 't Hooft coupling $\lambda$), then the ``current non-conservation relation" takes the form
\ie\label{jjj}
\partial^\mu J^{(s)}_{\mu\nu_1\cdots\nu_{s-1}} = {\widetilde f(\lambda)\over \sqrt{N}} \sum_{s_1+s_2<s} \partial^{n_1} J^{(s_1)} \partial^{n_2} J^{(s_2)} + {\widetilde g(\lambda)\over N} \sum_{s_1+s_2+s_3<s} \partial^{n_1} J^{(s_1)} \partial^{n_2} J^{(s_2)} \partial^{n_3} J^{(s_3)}.
\fe
Here we have omitted the nontrivial contraction of Lorentz indices on each term, which is essentially fixe by requiring the terms appearing on the RHS of (\ref{jjj}) to be conformal primaries at infinite $N$.\footnote{This is because $||P^\mu |J_{\mu\cdots}\rangle||^2\sim {\cal O}(N^{-1})$, which implies that for instance $K |JJ\rangle \sim {\cal O}(N^{-{1\over 2}})$, for the double trace operator $JJ$ appearing on the RHS of the current non-conservation relation.} At infinite $N$, while the current $J^{(s)}$ has twist $\Delta - s = 1$, its divergence has twist 3, and therefore can only be expressed in terms of a sum of products of two or three currents, but no more than three.

The current non-conservation relation (\ref{jjj}) is also closely related to the three-point functions \cite{Maldacena:2012sf}, as we will discuss later.

\section{Vasiliev perturbation theory}

In this section we formulate the perturbation theory of Vasiliev's system around the $AdS_4$ vacuum, and set up the computation of boundary correlation functions. See also \cite{Sezgin:2002ru} for an analysis of the perturbative expansion of Vasiliev's equations.

\subsection{Generalities}

In terms of $\widehat W$, $S$, and $B$, Vasiliev's equations can be written in the following form,
\ie\label{vaspert}
& D_0 \widehat W = - \widehat W* \widehat W,
\\
& d_Z \widehat W + D_0 S = -\widehat W*S-S*\widehat W,
\\
& d_Z S = e_*^{i\theta(B*K)}*B*Kdz^2 + e_*^{-i\theta(B*\overline K)}*B*\overline K d\bar z^2,
\\
& \widetilde D_0 B = - \widehat W*B + B*\pi(\widehat W),
\\
& d_Z B = - S*B+B*\pi(S).
\fe
where we wrote $e_*^{i\theta}$ for the $*$-exponential of the $*$-function $\theta$. In the explicit computation of the three-point function, $e_*^{i\theta}$ is simply taken to be the phase $e^{i\theta_0}$.

To solve Vasiliev's equations perturbatively, we begin with the first order fields $\widehat W^{(1)}, S^{(1)}, B^{(1)}$ which solve the linearized equations, plug them into the RHS of (\ref{vaspert}) and solve the second order fields, first $B^{(2)}$ from the last two equations of (\ref{vaspert}), then $S^{(2)}$ from the third equation of (\ref{vaspert}) and $\widehat W^{(2)}$ from the first two equations of (\ref{vaspert}), and so forth.

Suppose we have solved the order $k$ fields $B^{(k)}$, $S^{(k)}$, $\widehat W^{(k)}$, for $k\leq n-1$. We now want to solve for the $n$-th order field $B^{(n)}$, from the two equations
\ie\label{nbb}
&\widetilde D_0 B^{(n)} = \sum_{k=1}^{n-1} \left[ - \widehat W^{(k)}*B^{(n-k)} + B^{(n-k)}*\pi(\widehat W^{(k)}) \right],
\\
& d_Z B^{(n)} =  \sum_{k=1}^{n-1} \left[-S^{(k)}*B^{(n-k)} + B^{(n-k)}*\pi(S^{(k)})\right].
\fe
Let us define
\ie
C(x|Y) = \left. B\right|_{Z=0},
\fe
and split $B(x|Y,Z)$ into a $Z$-independent piece, which contains the higher spin Weyl curvatures, and a $Z$-dependent piece that is determined by lower order $S$ and $B$ fields, namely
\ie
B(x|Y,Z) = C(x|Y) + {\cal B}'(x|Y,Z),
\fe
where ${\cal B}'$ obeys ${\cal B}'|_{Z=0}=0$. 
The essentially nontrivial part is to solve for $C(x|Y)$ from the equation
\ie\label{cjeqn}
\widetilde D_0 C &=\left( \widetilde D_0 B \right) |_{Z=0} - \left( -W_0*{\cal B}' + {\cal B}'*\pi(W_0)\right) |_{Z=0}
\\
&= J^Y + J^Z \equiv J_\mu(x|Y) dx^\mu,
\fe
where
\ie\label{jy}
J^Y &= \left[-\widehat W*B + B*\pi(\widehat W)\right]_{Z=0}
\\
&= -\Omega*C + C*\pi(\Omega) + \left[-W'*C + C*\pi( W')\right]_{z=\bar z=0}+ \left[-\Omega*{\cal B}' + {\cal B}'*\pi( \Omega)\right]_{Z=0}
\\
&~~~ + \left[-W'*{\cal B}' + {\cal B}'*\pi( W')\right]_{Z=0},
\fe
and
\ie\label{jz}
J^Z &= \left( W_0*{\cal B}' - {\cal B}'*\pi(W_0)\right) |_{Z=0}.
\fe
At the $n$-th order, $J^Y$ is already expressed in terms of the lower order $\widehat W$ and $B$ fields, whereas $J^Z$ is given in terms of ${\cal B}'$ at the $n$-order, which is easily solved in terms of the lower order $S$ and $B$ fields by integrating the second equation of (\ref{nbb}).

To solve for $C$ from $J_\mu$, let us write (\ref{cjeqn}) as
\ie
\nabla_\mu^L C + \{ e_\mu, C\}_* = J_\mu,
\fe
where $\nabla_\mu^L$ is defined as in section 2.5.1, and $e_\mu(x|Y)$ is the vierbein of $AdS_4$ contracted with $y\bar y$ as before. We repeat the analysis of the linearized equation for $C$, now with a source $J_\mu$, turn the equation into second order form and solve for $C$ by inverting the second order kinetic operator.


Once we have solved $C(x|Y)$ and ${\cal B}'(x|Y,Z)$ at $n$-th order, we can then solve $S$, $W'$, and finally $\Omega$ at the same order, following the same analysis as the linearized equations earlier, only now with sources. In computing the three-point functions though, it suffices to compute $C(x|Y)$ at the second order, since the boundary correlators can be extracted entirely from the boundary limiting value of the higher spin Weyl curvatures \cite{Giombi:2009wh}.

\subsection{Bulk-to-boundary propagators}
In this section we collect the results of \cite{Giombi:2009wh} for the bulk-to-boundary propagators of the higher spin gauge fields and the Vasiliev's master fields. As usual, the bulk-to-boundary propagators are defined to be the solutions of the linearized equations with boundary conditions corresponding to the insertion of $\delta$-function source at the AdS boundary.
\subsubsection{The spin $s$ gauge field}
\label{btb-HS}
As explained in section \ref{LinEq-sec}, the linearized Vasiliev's equation for the physical degrees of freedom can be shown to be equivalent to the Fronsdal's equations for free higher spin fields in $AdS_4$. We will assume this equivalence and first derive the bulk-to-boundary propagator for the spin $s$ gauge fields in the Fronsdal's formulation. 

The gauge invariant Fronsdal's equations are written in terms of a rank $s$ symmetric tensor $\varphi_{\mu_1\cdots \mu_s}$ which is double traceless. Upon doing a partial gauge fixing to gauge away the trace part, and further choosing the transverse gauge condition $\nabla^\nu \varphi_{\nu \mu_1\cdots\mu_{s-1}}=0$, Fronsdal's equations in $AdS_{d+1}$ reduce to the wave equation
\begin{equation}
\begin{aligned}
&(\Box-m^2)\varphi_{\mu_1\cdots\mu_s}=0,\\
&m^2=(s-2)(d+s-3)-2\,.
\label{Fronsdal}
\end{aligned}
\end{equation}
Here $\Box=\nabla^{\mu}\nabla_{\mu}$. Note that $m^2$ does not of course represent a mass in the ordinary sense. This term comes from the curvature of AdS, and its value precisely corresponds to a massless spin $s$ field. 

A solution to the equation (\ref{Fronsdal}) has the boundary behavior as $z\to 0$,
\begin{equation}
\varphi_{i_1\cdots i_s}(\vec x,z)\sim z^\delta,~~~~(\delta+s)(\delta+s-d)-s=m^2.
\end{equation}
where the indices $i_k$ are along the boundary directions, running from $0$ to $d-1$.
From this we read off the dimension of the dual operator, a spin-$s$ current $J_{i_1\cdots i_s}$,
\begin{equation}
\Delta=d-\delta-s = {d\over 2} + \sqrt{m^2 +s+ \left({d\over 2}\right)^2} = d-2+s
\end{equation}
This scaling dimension also follows from the conformal algebra under the assumption that $J_{i_1\cdots i_s}$ is a conserved current
and a primary operator. In particular, in a free scalar field theory in $d$ dimensions, the currents of the
form $\phi \partial_{i_1}\cdots\partial_{i_s}\phi+\cdots$ have dimension $\Delta=d-2+s$. 

We will now specialize to $d=3$, i.e. $AdS_4$, and give the result for the boundary-to-bulk propagator which solves (\ref{Fronsdal}). We will not give the details of the calculation here, and refer the reader to \cite{Giombi:2009wh} for the complete derivation. To write the result in a compact way, it is convenient to introduce a generating function
\begin{equation}
\Phi_s(\vec{x},z|Y) = z^s\sum \varphi_{\mu_1\cdots\mu_s}(\vec{x},z) Y^{\mu_1}\cdots Y^{\mu_s} 
\end{equation}
where we have introduced an auxiliary variable $Y^{\mu}$. This play essentially the same role as the internal twistor variables $y_{\alpha}, \bar y_{\dot\alpha}$ in Vasiliev's formulation. Then the boundary-to-bulk propagator corresponding to a boundary spin $s$ source contracted with a null polarization vector $\vec{\varepsilon}$ is given by 
\ie
\Phi_s = \tilde N_s\left. e^{iux^\mu Y_\mu} {(\varepsilon\cdot (-i\vec\partial+u\vec Y))^{2s}\over (i\varepsilon\cdot \vec\partial)^s}\right|_{u^s} \left( z\over \vec{x}^2+z^2 \right)^{s+1}
\fe
for some normalization constant $\tilde N_s$. Here $|_{u^s}$ means to pick out the coefficient of
$u^s$ in a series expansion in $u$. Near the boundary $z\to 0$, one can show that the leading behavior of $\Phi_s(\vec{ x},z|Y)$ is given by
\ie
\Phi_s(x,z|Y) 
&\to \tilde N_s \pi^{3\over 2}{\Gamma(s-{1\over 2})(2s)!\over 2 (s!)^3} z^{2-s}(\varepsilon\cdot \vec Y)^s \delta^3(\vec x).
\fe
This is indeed the correct boundary behavior corresponding to the insertion of a 
higher spin source at the origin. We may of course insert it at any other position $\vec{x}_0$ by simply shifting $\vec{x}\rightarrow \vec{x}-\vec{x}_0$. One may fix the normalization by requiring that the coefficient of $z^{2-s}(\vec{\varepsilon}\cdot\vec Y)^s\delta^3(\vec x)$ is unity, so that the normalization constant $\tilde N_s$ is determined to be
\ie
\tilde N_s = {2\pi^{-{3\over 2}} (s!)^3\over \Gamma(s-{1\over 2})(2s)!}.
\fe

It is sometimes convenient to work in light cone coordinates on the boundary $\vec x = (x^+, x^-, x_\perp)$,
with $\vec x^2=x^+x^-+x_\perp^2$ and $\vec\varepsilon\cdot\vec \partial
=\partial_+$, i.e. $\varepsilon^+=1,\varepsilon^-=0$. Then a short calculation shows that we can then write the boundary-to-bulk propagator for
$\Phi_s$ simply as
\ie
\Phi_s = \tilde N_s {z^{s+1}\over (s!)^2(x^-)^s} \partial_+^{2s}  {(x^\mu Y_\mu)^s \over \vec x^2+z^2}.
\label{btb-phi-lCone}
\fe

\subsubsection{The master fields $C$ and $\Omega$}
Let us start by recalling the linearized equations (\ref{ceqn}) for the master field $C(x|Y)$. Writing the explicit vierbein and spin connection in Poincar\'e coordinates, they are
\ie 
\label{eomab}
dC -{dx^i\over 2z}\left[(\sigma^{iz})_\A{}^\B y^\A \partial_\B
+ (\sigma^{iz})_\da{}^\db \bar y^\da \partial_\db \right] C + {dx^\mu\over 2z}\sigma_\mu^{\A\db}
\left( y_\A \bar y_\db +\partial_\A\partial_\db \right) C =0.
\fe 

These may be written as a system of equations for the components $C^{(n,m)}$ of degree $n$ in $y$ and $m$ in $\bar y$. The equations only couple $C^{(n,m)}$'s with the same $|n-m|$ as explained earlier. The scalar field and its derivatives are contained in $C{(n,n)}$. By solving (\ref{eomab}) in the scalar sector we obtain the boundary-to-bulk propagator for the scalar component of the master field $C(x|Y)$. The answer for the $\Delta=1$ boundary condition is
\ie
\begin{aligned}
\label{eomac}
&C(x|y,\bar y) = K e^{-y(\sigma^z-2{\bf x}K)\bar y} = K e^{-y\Sigma \bar y},
\\
&K=\frac{z}{\vec{x}^2+z^2},
\end{aligned}
\fe
where we used the notation ${\bf x} \equiv x^\mu \sigma_\mu = x^i \sigma_i + z\sigma^z$. We have also defined 
$$\Sigma = \sigma^z-{2z\over x^2}{\bf x}.$$
With the alternative  $\Delta=2$ boundary condition, the boundary-to-bulk propagator for the scalar components of the master field turns out to be 
\ie
C^{\Delta=2}(x|y,\bar y) = K^2 (1-y\Sigma \bar y) e^{-y\Sigma\bar y}.
\label{C-prop-Del2}
\fe
 
Let us now consider the spin-$s$ components of $C$ (namely the components $C^{(n,m)}$ with $|n-m|=2s$). The solution of (\ref{eomab}) in the spin-$s$ sector takes the form
\ie
\label{eomad}
C = \frac{e^{i\theta_0}}{2}\, K e^{-y\Sigma \bar y} T(y)^{s}+c.c.,
\fe
where $T(y)$ is given by
\ie\label{eomae}
T(y) = {K^2\over z} y{\bf x}\vec{\varepsilon}\cdot\vec\sigma \sigma^z {\bf x} y,
\fe
for an arbitrary polarization vector $\vec\varepsilon$ along the 3-dimensional boundary. Relating $C$ to $\Omega$ via the linearized equation of motion (\ref{ewaa}), one may verify that this indeed is the master field corresponding to the boundary-to-bulk propagator for the spin-$s$ tensor
gauge field derived in section \ref{btb-HS}, dual to a boundary source contracted with polarization vector $\vec{\varepsilon}$. It will be useful in the following to trade a null polarization vector with a bispinor $\lambda_{\alpha}$. This may be introduced as 
$$2({\slash\!\!\!\varepsilon}\sigma_z)_{\A\B} = \lambda_\A\lambda_\B,\qquad  2({\slash\!\!\!\varepsilon}\sigma_z)_{\da\db}=\bar\lambda_\da \bar\lambda_\db$$
with $\bar\lambda=\sigma^z\lambda$ (the factor of $2$ here is just our choice of convention). Then, the spin $s$ bulk-to-boundary propagator may be written as
\begin{eqnarray}
C(x|y,\bar y)
&=& \frac{e^{i\theta_0}}{2^{s+1}}\, \frac{z^{s+1}}{\left(\vec{x}^2+z^2\right)^{2s+1}}e^{-y\Sigma \bar y} \left( y{\bf x}\sigma^z\lambda\right)^{2s}+ c.c.
\label{C-prop}
\end{eqnarray}

Let us finally give the bulk-to-boundary propagator for the master field $\Omega$, which (after gauge fixing) is directly related to the rank $s$ symmetric tensor higher spin field. We choose the following normalization convention for $\Omega^{(s-1,s-1)}$ in terms of $\Phi_s$
\ie
\label{omnorm}
\Omega_{\A\db}^{(s-1,s-1)} = {(s!)^2\over 2\tilde N_s (2s)!}{1\over s z}\partial_\A\partial_\db \Phi_s(x|Y^\mu = y\sigma^\mu\bar y)
\fe  
In terms of the notation introduced in (\ref{Omega-shorthand}), and using the explicit propagator (\ref{btb-phi-lCone}), this may be written as (here we specialize to light-cone coordinates and $\varepsilon^+=1$)
\begin{equation} 
\Omega_{++}^0 = {(s!)^2\over 2\tilde N_s (2s)!}{s\over z}\Phi_s(x| y\sigma^\mu\bar y)= {s z^{s}\over 2 (2s)! (x^-)^s}\partial_+^{2s}{(y{\bf x}\bar y)^s\over x^2}.
\end{equation}
The higher components $\Omega^n_{++}, n=1,\cdots,s-1$ can be obtained through the relations (\ref{omegas}), and one finds
\ie
\Omega_{++}^n &= {(s-n)!\over s(s+n-1)!} z^{-n}(z^2 y{\slash\!\!\!\partial}\partial_{\bar y})^n \Omega_{++}^0
\\
&= {2^{-n-2}\over (2s-1)!} {z^s\over (x^-)^{s+n}} (y{\bf x}\sigma^{-z}{\bf x} y)^n
\partial_+^{2s} {(y{\bf x}\bar y)^{s-n}\over x^2}\,.
\fe
On the other hand, the bulk-to-boundary propagator for $\Omega_{-+}^n$ can be shown to vanish identically upon using (\ref{omegas}), see \cite{Giombi:2009wh} for details. Therefore we can recover the full bulk-to-boundary propagator for the master field $\Omega$ from $\Omega^n_{++}$.

An important property of both propagators of $C$ and $\Omega$ is that the divergence with respect to the position of the boundary source vanishes, namely 
\begin{equation}
\begin{aligned}
&\partial_{\lambda}{\slash\!\!\!\partial}_{x_0}\partial_{\lambda}C(\vec{x}-\vec{x}_0,z|y,\bar y)=0\\
&\partial_{\lambda}{\slash\!\!\!\partial}_{x_0}\partial_{\lambda}\Omega(\vec{x}-\vec{x}_0,z|y,\bar y)=0\,.
\end{aligned}
\end{equation}
Here $\lambda$ is the spinor polarization of the boundary sources, as introduced earlier.
This guarantees that the holographic correlation functions computed from the bulk Vasiliev's theory will obey current conservation, modulo possible contact terms which can arise with higher spin symmetry breaking boundary conditions and which are responsible for violation of current conservation within correlation functions.

\subsection{Boundary correlators}

\begin{figure}
\begin{center}
\includegraphics[width=80mm]{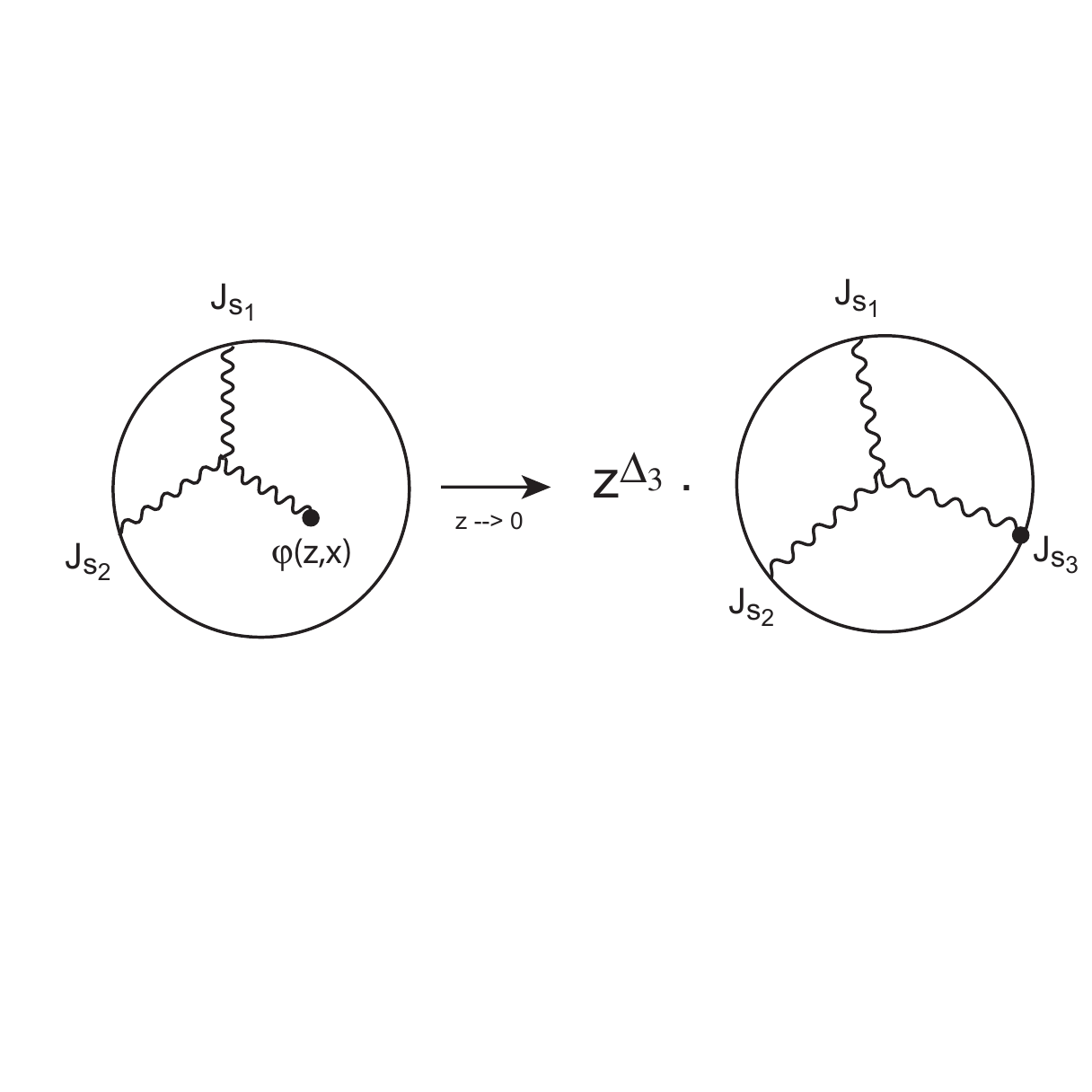}
\vskip -2.5cm
\parbox{13cm}{
\caption{Extracting the three-point function from the boundary behavior of the second order field sourced by $J_{s_1}$, $J_{s_2}$.}}
\end{center}
\label{Witten-diag}
\end{figure}

Usually, holographic correlation functions are computed in perturbation theory by Witten diagrams \cite{Witten:1998qj}, where boundary-to-bulk propagators are sewed with bulk propagators and vertices, integrated over AdS spacetime. In principle, one could try to recover the bulk action for Vasiliev's system order by order in the fields, starting from the nonlinear equations of motion expressed in terms of the metric-like higher spin fields, and then extract the interaction vertices from such an action. In practice, such an approach appears to be extremely messy. However, at least for tree level correlators, an explicit form of the bulk action is not necessary. The boundary correlators can equivalently be computed by directly using the equation of motion, as follows. Suppose we want to compute a boundary correlator $\langle {\cal O}_1(\vec x_1)\cdots {\cal O}_n(\vec x_n)\rangle$ at tree level. We can treat $n-1$ of these operators, say ${\cal O}_1,\cdots,{\cal O}_{n-1}$, as boundary sources, and compute the bulk field $\varphi_n(\vec x,z)$ dual to ${\cal O}_n$, sourced by these boundary operators, by solving the bulk equation of motion to the $(n-1)$-th order. The $n$-point function is then proportional to the boundary value of $\varphi_n(\vec x,z)$, after an appropriate factor $z^\delta$ is stripped off in the $z\to 0$ limit, see Fig. \ref{Witten-diag}. This procedure by itself does not quite fix the normalization of the $n$-point function, since without a bulk action the normalization of the propagator is arbitrary. The relative normalization can be fixed by comparing different channels related by crossing. Namely, if we choose any other set of $n-1$ operators out of ${\cal O}_1,\cdots,{\cal O}_n$ as the boundary source, and compute the correlator from the boundary limiting value of the field dual to the remaining operator, we must arrive at the same $n$-point function modulo a normalization factor that can be absorbed into the propagators. By comparing different crossing channels, one can fix the normalization of all $n$-point correlators up to a single overall factor, which is related to the overall coupling constant $g$ of Vasiliev theory. In fact, since $g$ drops out of the classical equation of motion, it must be put in by hand for each bulk vertex. Once a normalization convention for the bulk coupling $g$ is given, its relation with the boundary CFT, of the form $g\sim 1/\sqrt{N}$ for large $N$ vector models, can be determined for instance by comparing the normalization of any three-point function.

For the purpose of extracting boundary correlators, we can focus on the self-dual part of the spin-$s$ Weyl curvature, $C^{(2s,0)}$. This is because in the boundary limit $z\to 0$, $C^{(2s,0)}$ is proportional to the metric-like rank-$s$ symmetric traceless tensor field up to a power of $z$. It suffices to consider the following components of the $C$ equation of motion with sources,
\ie
&
\nabla_\mu^L C^{(2s,0)} + 2 e_\mu^{\A\db} \partial_{y^\A}\partial_{\bar y^\db} C^{(2s+1,1)} = J_\mu^{(2s,0)},
\\
&
\nabla_\mu^L C^{(2s+1,1)} + 2e_\mu C^{(2s,0)} + 2 e_\mu^{\A\db} \partial_{y^\A}\partial_{\bar y^\db} C^{(2s+2,2)} = J_\mu^{(2s+1,1)}.
\fe
By the same manipulation as in the analysis of the linearized equations, we can eliminate $C^{(2s+2,2)}$ and $C^{(2s+1,1)}$, and write a second order equation for $C^{(2s,0)}$ by itself,
\ie\label{cceqn}
&e^{\nu \C\dd}\partial_{y^\C} \partial_{\bar y^\dd}\nabla_\nu^L \left[ e^{\mu}_{\A\db} y^\A \bar y^\db\nabla_\mu^L C^{(2s,0)}\right] +{2s+1\over 2}  e^{\mu \A\db}\partial_{y^\A} \partial_{\bar y^\db}\left[ e_\mu C^{(2s,0)}\right] 
\\
&= e^{\nu \C\dd}\partial_{y^\C} \partial_{\bar y^\dd}\nabla_\nu^L \left[e^{\mu}_{\A\db} y^\A \bar y^\db J_\mu^{(2s,0)}\right] + {2s+1\over 4} e^{\mu \A\db}\partial_{y^\A} \partial_{\bar y^\db}J_\mu^{(2s+1,1)} .
\fe
Note that the RHS depends on $J_\mu$ only through the components $J^{(2s,0)}$ and $e^{\mu \A\db}\partial_{y^\A} \partial_{\bar y^\db}J_\mu^{(2s+1,1)}$.

To proceed further, it is convenient to work with the explicit expressions for $AdS_4$ vierbein and spin connection in Poincar\'e coordinates.
The equation (\ref{cceqn}) can be written in the form
\ie\label{cjj}
&\left[ z^2 \partial^\mu\partial_\mu - 2z \partial_z + z (y \sigma^{zi}\partial_y ) \partial_i -(s-2)(s+1)  \right] C^{(2s,0)} = -{2z\over s+1} {\cal J}(y),
\fe
where $z$ is the Poincar\'e radial coordinate, not to be confused with the $Z$-twistor variables which do not appear here. The source $J(y)$ is given by 
\ie\label{jsou}
{\cal J}(y)&=\partial^\A \partial^\db \nabla_{\A\db} z J_{\C\dd}^{(2s,0)} y^\C \bar y^\dd - {2s+1\over 2}\partial^\A\partial^\db J_{\A\db}^{(2s+1,1)}  
\\
&= -{z\over 2}\partial_y ({\slash\!\!\!\partial}\,-{s+2\over z}\sigma^z) {\slash\!\!\! J}^{(2s,0)} y - {2s+1\over 2}\partial^\A\partial^\db J_{\A\db}^{(2s+1,1)} .
\fe
(\ref{cjj}) can now be solved by integrating the source with a propagator ${\cal K}$.
The behavior of the outcoming spin-$s$ field near the boundary is given by
\ie
&C^{(2s,0)}(\vec x,z\to 0|y)  \to  z^{s+1} \int {dz_0 d^3\vec x_0\over z_0^4} {\cal K}(\vec x-\vec x_0, z_0|y,\partial_{y_0}) \left[ -{2z_0\over s+1} {\cal J}(\vec x_0,z_0|y_0) \right] .
\fe
The boundary correlation function involving a spin-$s$ current $J^{(s)}$ can be extracted from the helicity-$s$
part of $C^{(2s,0)}(\vec x,z\to 0|y)$. It then suffices to take
the helicity-$s$ part of the propagator ${\cal K}$, denoted by ${\cal K}_{(s)}$, and given explicitly by
\ie
&{\cal K}_{(s)}(\vec x,z|y,\lambda) =
{2^{-2s} z^{2-s} }\int_0^\infty dt
{(1+t)^{-2s}  } \\
&~~~\times \left.\left\{ {(y \sigma^z{\slash\!\!\!\partial}\lambda)^{2s}\over (2s)!} \left[-{\Gamma(2-2s)(x^2)^{s-1}\over 2\pi^2|\vec x|}{
\sin\left(2(s-1)\arctan{|\vec x|\over z} \right) } \right] \right\}\right|_{z\to (2t+1)z}
\fe
Away from $\vec x=0$, ${\cal K}_{(s)}$ has an expansion around $z=0$ of the form
\ie\label{kzser}
{\cal K}_{(s)}(\vec x,z|y,\lambda) = z^{2-s}\sum_{n=0}^\infty a_n^{(s)}(\vec x|y,\lambda) z^n + z^{s+1}\log(z)
\sum_{n=0}^\infty b_n^{(s)}(\vec x|y,\lambda).
\fe
Importantly, $b_0^{(s)}(\vec x|y,\lambda)$ is given by
\ie
b_0^{(s)}(\vec x|y,\lambda) = {\cal N}_s{(y{\bf \hat x}\sigma^z \lambda)^{2s}\over (2s)!(x^2)^{2s+1}},
~~~~~{\cal N}_s = {2^{2s-1}s\over \pi^2}.
\fe
where ${\bf \hat x}\equiv \vec x\cdot\vec \sigma$. The other helicity components ${\cal K}_{(m)}$, for $m<s$,
when expanded near $z=0$, will only have the first branch of (\ref{kzser}) and not the second branch with the $\log(z)$ factor.

The scalar field is a special case. For $s=0$, the dual operator can have dimension
$\Delta=1$ or $\Delta=2$. ${\cal K}_{(0)}$ is simply given by
\ie\label{ksimple}
&{\cal K}_{(0)}^{\Delta=1}(\vec x,z) = {1\over 2\pi^2}{z\over \vec x^2+z^2},\\
&{\cal K}_{(0)}^{\Delta=2}(\vec x,z) = {1\over \pi^2}{z^2\over (\vec x^2+z^2)^2},\\
\fe

The boundary $n$-point function of the spin-$s$ current $J^{(s)}$, of the form $\langle J^{(s)}(\vec x|y)\cdots\rangle$ where $J^{(s)}(\vec x|y) = J^{(s)}_{\A_1\cdots\A_{2s}}(\vec x) y^{\A_1}\cdots y^{\A_{2s}}$, can be computed by regarding the other $n-1$ operator insertions as boundary sources, solving for the $(n-1)$-th order bulk spin-$s$ field, and extracting the boundary limiting value of the spin-$s$ Weyl curvature tensor contained in $C^{(2s,0)}(\vec x,z|y)$,
\ie
\lim_{z\to 0} z^{-s-1} C_{h=s}^{(2s,0)}(\vec x,z|y) = \int {dz_0 d^3\vec x_0\over z_0^4} {\cal K}_{(s)}(\vec x-\vec x_0,z_0|y, \partial_{y_0}) \left[ - {2z_0\over s+1} {\cal J}(\vec x_0,z_0|y_0) \right].
\fe

\section{Holographic three-point functions}

In this section we carry out the explicit computation of three-point functions of boundary currents in non-minimal bosonic Vasiliev theory. We first present the computation in the physical spacetime, by deriving the source ${\cal J}(x|y)$ at the second order from the linearized fields sourced by two boundary currents, and then evaluate the boundary limiting value of the bulk field sourced by ${\cal J}(x|y)$. We will restrict ourselves to the case where one of the three currents is a scalar, whereas the other two currents are of general spins. It will turn out that only $J^\Omega$ contributes to ${\cal J}(x|y)$, which makes the computation particularly simple. For the $s-s'-0$ correlator, we will find that the result is a linear combination of a parity even structure and a parity odd structure, with coefficient $\cos\theta_0$ and $\sin\theta_0$, where $\theta_0$ is the parity breaking phase in Vasiliev theory. The result is precisely consistent with the general structure constrained by ``slightly broken" higher spin symmetry \cite{Maldacena:2012sf}. While the computation of the parity invariant case was performed in \cite{Giombi:2009wh}, the computation of parity odd contributions presented here is new.

The physical spacetime approach can in principle be used to compute the more general three-point functions, with all three currents of nonzero spins, but in practice it gets very cumbersome. Curiously, Vasiliev's system allows for a formal (large) gauge transformation (which does not preserve the $AdS$ boundary condition) that gauges away the explicit spacetime dependence in the master fields entirely. This is sometimes referred to as the ``$W=0$ gauge", or the gauge function method, see e.g. \cite{Vasiliev:1990bu, Bolotin:1999fa, Bekaert:2005vh, Sezgin:2005pv, Iazeolla:2007wt, Iazeolla:2011cb}. Working in the $W=0$ gauge allows one to solve the master fields as functions of the twistor variables $(Y,Z)$ only, which is technically very simple. In order to extract the boundary correlators, one must first transform the solutions back to the physical gauge, where spacetime dependence is restored, and then take the boundary limiting value of the higher spin field \cite{Giombi:2010vg}. We will carry out this computation explicitly in the parity invariant A-type and B-type non-minimal bosonic theories. While this approach is in principle straightforward, there are two potentially important subtleties that are glossed over in our computation. The first one is that the integral representation of the star product between two fields will be singular, and a contour prescription is introduced to regularize the integral. The second subtlety is that there is some ambiguity in transforming the fields from the $W=0$ gauge back to the physical gauge, and this gauge ambiguity has not been fixed properly though appears to be absent in the computation of three-point functions in the parity invariant theory. It appears, on the other hand, that the parity odd contributions in the case of generic parity breaking phase $\theta_0$ are closely related to the gauge ambiguity. We hope to report on a proper treatment of the gauge ambiguity in the near future.

\subsection{Physical space-time approach}

To proceed, we shall compute $J^Y$ (\ref{jy}) and $J^Z$ (\ref{jz}) at the second order in fields. Making use of the facts that linearized master field $B^{(1)}=C^{(1)}$ is independent of $Z$, and that the $Z$-dependent part of the second order master field ${\cal B}'^{(2)}$ is the sum of a function of $z$ only and another function of $\bar z$ only, with no mixed $(z,\bar z)$ dependence, we can write
\ie
J^{Y(2)} &= -\Omega^{(1)}*C^{(1)} + C^{(1)}*\pi(\Omega^{(1)}) + \left[-W'^{(1)}*C^{(1)} + C^{(1)}*\pi( W'^{(1)})\right]_{z=\bar z=0} 
\\
& = J^{\Omega(2)} + J'^{(2)},
\\
J^{Z(2)} &= -\epsilon^{\A\B} \left( \partial_{y^\A} W_0*\partial_{z^\B}{\cal B}'^{(2)} |_{z=0} -\partial_{z^\B} {\cal B}'^{(2)}|_{z=\bar z=0} *\partial_{y^\A}\pi(W_0)\right) + c.c.
\\
&= -\epsilon^{\A\B} \left[ \partial_{y^\A} W_0*(S_\B^{(1)}*C^{(1)} - C^{(1)}*\pi(S_\B^{(1)}))|_{z=0} \right.
\\
&\left.~~~~~~~~~ - (S_\B^{(1)}*C^{(1)} - C^{(1)}*\pi(S_\B^{(1)}))|_{z=0} *\partial_{y^\A}\pi(W_0)\right]  + c.c.
\fe
At this point, we need to make use of the explicit expressions of $S^{(1)}$ and $W'^{(1)}$ in terms of $C^{(1)}$,
\ie
& S^{(1)}_\A = -e^{i\theta_0}  z_\A \int_0^1 dt \,t\, C^{(1)}(-tz,\bar y) e^{tzy},
\\
& W'^{(1)} = 2 e^{i\theta_0} z^\A \int_0^1 dt\,(1-t) \left[ (e_0)_\A{}^\db \partial_{\bar y^\db} + (\omega_0)_{\A\B} tz^\B \right] C(x|-tz,\bar y) e^{tzy} + c.c.
\fe
We have already seen from (\ref{jsou}) that in order to solve for the spin-$s$ self-dual Weyl curvature contained in $C^{(2s,0)}$, we only need to know $J^Y|_{\bar y=0}$, $J^Z|_{\bar y=0}$, and $(\sigma^\mu)^{\A\db}\partial_{y^\A}\partial_{\bar y^\db} J^Y_\mu|_{\bar y=0}$, $(\sigma^\mu)^{\A\db}\partial_{y^\A}\partial_{\bar y^\db} J^Z_\mu|_{\bar y=0}$. 

Let us examine the possible powers of $y$ appearing in $J'^{(2)}|_{\bar y=0}$, $J^{Z(2)}|_{\bar y=0}$, and $(\sigma^\mu)^{\A\db}\partial_{y^\A}\partial_{\bar y^\db} J'^{(2)}_\mu|_{\bar y=0}$, $(\sigma^\mu)^{\A\db}\partial_{y^\A}\partial_{\bar y^\db} J^{Z(2)}_\mu|_{\bar y=0}$. They are expressed in terms of the star product of $W'^{(1)}, S^{(1)}$ with $C^{(1)}$. Consider the contribution from the spin-$s$ component of $W'^{(1)}, S^{(1)}$ and the spin-$s'$ component of $C^{(1)}$. We have, schematically,
\ie
& S^{(1)}_\A \supset z_\A z^{2s+n} \bar y^n (zy)^m,~~~~z_\A z^{n} \bar y^{2s+n} (zy)^m,
\\
& W'^{(1)} \supset z^{2s+n+2} \bar y^n (zy)^m,~~~~z^{n+2} \bar y^{2s+n} (zy)^m,~~~~\bar z^{2s+n+2} y^n (\bar z\bar y)^m,~~~~\bar z^{n+2} y^{2s+n} (\bar z\bar y)^m,
\\
& C^{(1)}\supset y^{2s'+k} \bar y^k,~~~ y^k \bar y^{2s'+k},
\fe
where $n,m,k$ are non-negative integers. We see that
\ie
&J'^{(2)}|_{\bar y=0}\sim W'^{(1)}* C^{(1)}|_{\bar y=0} \supset y^m\partial_y^{2s+k+2+m} y^{2s'+k},~~~ y^m \partial_y^{2s'-2s+k+2+m} y^k,~~~ y^m \partial_y^{n+2+m} y^{2s+2s'+n},
\\
&~~~~ y^{2s'+2s+2n+2-2\ell} ~(0\leq \ell\leq n,{\rm min}(2s+n,2s'+n+2)),
\\
&~~~~ y^{2s-2s'+2n+2-2\ell}~(0\leq \ell\leq {\rm min}(n,2s-2s'+n+2),{\rm min}(2s+n,n+2-2s')).
\fe
Here $\ell\geq 0$ is the number of pairs of $y$ that are contracted upon taking the star product. The first three terms have degrees in $\pm 2s\pm 2s'-2$ in $y$, and will cancel in their contribution to $J'^{(2s'',0)}$. The last two terms could contribute if $s\pm s'+n+1-\ell = s''$. Note that due to the constraints on the range of $\ell$, $J'^{(2)}|_{\bar y=0}$ contributes only when $s''> |s-s'|$.

By the same straightforward though tedious analysis, one can show that $J^{Z(2)}|_{\bar y=0}$, $(\sigma^\mu)^{\A\db}\partial_{y^\A}\partial_{\bar y^\db} J'^{(2)}_\mu|_{\bar y=0}$, and $(\sigma^\mu)^{\A\db}\partial_{y^\A}\partial_{\bar y^\db} J^{Z(2)}_\mu|_{\bar y=0}$ also contribute only if $s''>|s-s'|$.
If $s''\leq |s-s'|$, then the only nonzero contribution comes from
\ie
J^{\Omega (2)}=  -\Omega^{(1)}*C^{(1)} + C^{(1)}*\pi(\Omega^{(1)}).
\fe
In the next section we will compute the contribution from $J^\Omega$ in the case $s''=0$, i.e. the outgoing second order field is taken to be the bulk scalar. The case $s=s'$ is special and in fact singular: naively the contribution from $J^\Omega$ vanishes, but if we take the formula for general $s\not=s'$ and analytically continue to $s=s'$, a nonzero answer is recovered (and will agree with the expected answer in the dual boundary CFT).

\subsection{The spin $s$-$s'$-$0$ correlator}
\label{sspz}

In computing the three-point function of currents of spin $s$, $s'$, and $0$, we can treat either $J^{(s)}$ and $J^{(s')}$ as sources, and extract the correlation function from the boundary limiting value of the scalar field, or treat $J^{(s)}$ and the scalar operator as sources, and extract from the boundary limiting value of the spin-$s'$ field. Since we have thus far not specified the normalization of the boundary-to-bulk propagators, it is useful to compare the computations in different channels related by crossing in order to fix the relative normalizations. This is done in \cite{Giombi:2009wh} and we will not repeat the analysis here. Here we redo the computation of the bulk scalar sourced by $J^{(s)}$ and $J^{(s')}$ in the physical gauge, but now keeping track of all position and polarization dependence and allow for a general parity breaking phase $\theta_0$, which will allow us to extract the parity odd contribution as well.
 
We will assume for the moment that $s>s'$. Without loss of generality, we will assume that $J^{(s)}$ is inserted at the point $\vec x_1=0$ on the boundary, and $J^{(s')}$ is inserted at $\vec x_2$. We will write $\vec{\tilde x}\equiv \vec x-\vec x_2$ below, and $x^2 = \vec x^2 + z^2$, $\tilde x^2 = \vec{\tilde x}^2 + z^2$, etc. The polarization spinors for $J^{(s)}$ and $J^{(s')}$ are denoted $\lambda$ and $\lambda'$ respectively. The scalar component of the source ${\cal J}$ in (\ref{cceqn}) is computed as in equation (4.82) of \cite{Giombi:2009wh}, ${\cal J}={\cal J}^+ + {\cal J}^-$. When $s+s'$ is even, we have
\ie\label{jplus}
&{\cal J}^+(\vec x,z) = e^{i\theta_0} 2^{-s-s'-2} s  \int d^4u d^4v \cosh(uv+\bar u\bar v) \left[ {z\over s^2-s'^2} v \left( {\slash\!\!\!\partial} - {2\over z}\sigma^z \right)\bar v - 1\right]
\\
&~~~ \times  {z^{s+s'+1}\over (x^2)^{2s+1}(\tilde x^2)^{2s'+1}} e^{-v \widetilde\Sigma \bar v} (v {\bf\tilde x} \sigma^z\lambda')^{2s'} (u {\bf x}\sigma^z\lambda)^{s+s'} (\bar u {\bf x} \lambda)^{s-s'} 
\\
&= -e^{i\theta_0}{2^{-s-s'-2} s} \left\{ {z\over s+s'}\left[ \partial_u \left( {\slash\!\!\!\partial} - {2\over z}\sigma^z \right) {\bf x} \lambda \right] {(\partial_u \widetilde\Sigma {\bf x}\lambda)^{s-s'-1}} + {(\partial_u \widetilde\Sigma {\bf x}\lambda)^{s-s'}} \right\} 
\\
&~~~ \times  {(s+s')!\over (s-s')!} {z^{s+s'+1}\over (x^2)^{2s+1}(\tilde x^2)^{2s'+1}}(\lambda \sigma^z {\bf x} {\bf\tilde x} \sigma^z \lambda')^{2s'} (u {\bf x}\sigma^z\lambda)^{s-s'} .
\fe
Here the integral over $(u,\bar u,v,\bar v)$ comes from taking the star product; it is understood that the integration contours in $(u^\A,v^\A)$ are along $e^{i\pi/4}\mathbb{R}$, the contours in $(\bar u^\da,\bar v^\da)$ are along $e^{-i\pi/4}\mathbb{R}$, and the integration measure is normalized so that $\int d^2u\, e^{u^\A v_\A} = \delta^2(v)$, etc. ${\cal J}^-$ is given by the same expression with the substitution $\theta_0 \to -\theta_0$, $\vec x\to -\vec x$, $\vec{\tilde x}\to -\vec{\tilde x}$. When $s+s'$ is odd, ${\cal J}$ vanishes identically in Vasiliev theory without Chan-Paton factors, and the $s$-$s'$-$0$ correlator vanishes. With Chan-Paton factors, the $s$-$s'$-$0$ correlator is obtained from a similar expression for ${\cal J}^+$ as (\ref{jplus}), multiplied by the structure constant of the non-abelian gauge group, and ${\cal J}^-$ is given by a similar expression with an extra sign $(-)^{s+s'}$. In either case, we immediately learn the dependence of the $s$-$s'$-$0$ correlator on the parity breaking phase $\theta_0$: the parity even contribution comes with a factor $\cos\theta_0$, whereas the parity odd contribution comes with a factor $\sin\theta_0$. We now examine these structures more explicitly.

The $s$-$s'$-$0$ three point function, obtained by integrating ${\cal J}(\vec x,z)$ with the scalar boundary-to-bulk propagator, is (assuming $s+s'$ is even)
\ie\label{proa}
&e^{i\theta_0}{2^{-s-s'-1} s\over s+1} \int {dz d^3\vec x\over z^4} {z^2\over (\vec x-\vec x')^2+z^2} \left\{ {z\over s+s'}\left[ \partial_u \left( {\slash\!\!\!\partial} - {2\over z}\sigma^z \right) {\bf x} \lambda \right] {(\partial_u \widetilde\Sigma {\bf x}\lambda)^{s-s'-1}} + {(\partial_u \widetilde\Sigma {\bf x}\lambda)^{s-s'}} \right\} 
\\
& \times  {(s+s')!\over (s-s')!} {z^{s+s'+1}\over (x^2)^{2s+1}(\tilde x^2)^{2s'+1}}(\lambda\sigma^z{\bf x} {\bf\tilde x}\sigma^z \lambda')^{2s'} (u {\bf x}\sigma^z\lambda)^{s-s'} + (\theta_0 \to -\theta_0,\vec x\to -\vec x,\vec{\tilde x}\to -\vec{\tilde x},\vec x'\to -\vec x')
\\
& = e^{i\theta_0}{2^{-2s'-1} s (s+s')!\over s+1} \int {dz d^3\vec x} {z^{2s - 1}\over (x^2)^{2s+1}(\tilde x^2)^{s+s'} [(\vec x-\vec x')^2+z^2]}(\lambda\sigma^z{\bf x} {\bf\tilde x}\sigma^z \lambda')^{2s'} 
\\
& ~~~\times \left\{ {1\over s+s'} {\lambda\sigma^z{\bf x}({\bf x}-{\bf x}'){\bf x}' \lambda \over (\vec x-\vec x')^2+z^2} {\left(  \lambda \sigma^z {\bf x}{\bf \tilde x}({\bf\tilde x}-{\bf x})\lambda \right)^{s-s'-1}} + {\left( \lambda \sigma^z {\bf x}{\bf \tilde x}({\bf\tilde x}-{\bf x})\lambda \right)^{s-s'}\over \tilde x^2} \right\}
\\
&~~~  + (\theta_0 \to -\theta_0,\vec x\to -\vec x,\vec{\tilde x}\to -\vec{\tilde x},\vec x'\to -\vec x').
\fe
In above we have integrated by part and used the definition $\widetilde\Sigma = \sigma^z - {2z\over \tilde x^2}{\bf\tilde x}$.
Now shifting the positions of $J^{(s)}, J^{(s')}, J^{(0)}$ to $\vec x_1, \vec x_2, \vec x_3$ respectively, we can rewrite (\ref{proa}) in the form
\ie\label{intres}
&e^{i\theta_0}{2^{-2s'-1} s (s+s')!\over s+1} \int {dz d^3\vec x}{z^{2s-1} (\lambda \sigma^z ({\bf x}-{\bf x}_1) ({\bf x} - {\bf x}_2)\sigma^z  \lambda')^{2s'}  \over [(\vec x-\vec x_1)^2+z^2]^{2s+1}[(\vec x-\vec x_2)^2+z^2]^{s+s'} [ (\vec x-\vec x_3)^2+z^2]}
\\
&~~~ \times \left\{ -{1\over s+s'} {(\lambda \sigma^z ({\bf x}-{\bf x}_1)({\bf x}-{\bf x}_3)  {\bf x}_{13} \lambda) (\lambda\sigma^z ({\bf x}-{\bf x}_1) ({\bf x}-{\bf x}_2){\bf x}_{12} \lambda)^{s-s'-1} \over (\vec x-\vec x_3)^2+z^2}  \right.
\\
&\left.~~~ + {(\lambda \sigma^z ({\bf x}-{\bf x}_1) ({\bf x}-{\bf x}_2) {\bf x}_{12} \lambda)^{s-s'} \over (\vec x-\vec x_2)^2+z^2} \right\}  
+ (\theta_0 \to -\theta_0, \vec x_i \to -\vec x_i )
\\
&= e^{i\theta_0}{2^{-2s'-1} s (s+s')!\over s+1} \int {dz_0 d^3\vec x_0}{z_0^{2s-1} (\lambda \sigma^z {\bf x}_{01} {\bf x}_{02}\sigma^z  \lambda')^{2s'}  \over (x_{01}^2)^{2s+1}(x_{02}^2)^{2s'+1}x_{03}^2}
\\
&~~~ \times \left\{ -{1\over s+s'} {(\lambda \sigma^z {\bf x}_{01} {\bf \check x}_{03}  {\bf x}_{13} \lambda) (\lambda\sigma^z {\bf x}_{01}  {\bf \check x}_{02}{\bf x}_{12} \lambda)^{s-s'-1}}  + (\lambda \sigma^z {\bf x}_{01}  {\bf\check x}_{02} {\bf x}_{12} \lambda)^{s-s'}  \right\}  
\\
&~~~
+ (\theta_0 \to -\theta_0, \vec x_i \to -\vec x_i ).
\fe
where we have defined $\check x^\mu = x^\mu/x^2$, and ${\bf\check x} = \check x^\mu \sigma_\mu$.

Let us illustrate how the integral (\ref{intres}) is evaluated in the simplest nontrivial case, 1-0-0. As already mentioned, this correlator is non-vanishing only when nontrivial Chan-Paton factors are introduced, and the result takes the same form as given by (\ref{intres}) (multiplied by appropriate group theory factor). The contribution proportional to $e^{i\theta_0}$ involves the integral
\ie
& \int {dz_0 d^3\vec x_0}{z_0 \over (x_{01}^2)^{3} x_{02}^2 x_{03}^2}
\big[ - (\lambda \sigma^z {\bf x}_{01} {\bf \check x}_{03}  {\bf x}_{13} \lambda)  + (\lambda \sigma^z {\bf x}_{01}  {\bf\check x}_{02} {\bf x}_{12} \lambda) \big]
\\
&= \left[\lambda \sigma^z{\bf x}_{12}\lambda + {1\over 4} x_{12}^2 (\lambda \sigma^z {\slash\!\!\!\partial}_{x_1} \lambda) \right] \int {dz_0 d^3\vec x_0}{z_0 \over (x_{01}^2)^2 (x_{02}^2)^2 x_{03}^2} - (2\leftrightarrow 3)
\\
&=\left[\lambda \sigma^z{\bf x}_{12}\lambda + {1\over 4} x_{12}^2 (\lambda \sigma^z {\slash\!\!\!\partial}_{x_1} \lambda) \right]  {\pi^3\over 4 |x_{12}|^3 |x_{13}||x_{23}|} - (2\leftrightarrow3)
\\
&= - {\pi^3\over 8} {\lambda \sigma^z {\bf\check x}_{23}\lambda \over |x_{12}| |x_{13}| |x_{23}|}.
\fe
This structure is, of course, entirely fixed by conformal symmetry. Note that the two terms in the bracket $\{\cdots\}$ in (\ref{intres}) in fact give exactly the same contribution.

While we do not know an explicit closed form expression of the result of this integral as a function of $\vec x_i$ and $\lambda,\lambda'$ for general spins $s$ and $s'$, it is easy to see that the integral expression is conformally invariant.\footnote{The only nontrivial part is to verify that under the inversion $x^\mu\to \check x^\mu$, $\lambda \to {\bf\check x}\lambda$, the integral transforms to itself with the extra factor $x_1^2 x_2^2 x_3^2$. This is easily seen by noting that ${\bf x}_{ij}$ transforms to $-{\bf \check x}_i{\bf x}_{ij}{\bf \check x}_j$ under inversion.} Further, since our computation of $s$-$s'$-$0$ amounts to integrating $J^\Omega$ with a kernel, and $J^\Omega$, obtained by taking the star product of the boundary-to-bulk propagators of $\Omega(x|Y)$ and $C(x|Y)$, obeys conservation with respect to the spin-$s$ and spin-$s'$ sources up to possible contact terms, the integral (\ref{intres}) should also obey current conservation up to the contribution from these contact terms. The contact terms in the divergence with respect to the spin-$s$ source may give rise to a non-vanishing divergence of the three-point function, that takes a factorized form \cite{Giombi:2011ya}, which is accounted for by the current non-conservation relation that relates the divergence of a spin-$s$ current to the product of currents of lower spins, including spin $s'$ and spin $0$. Conformal invariance together with the ``almost" conservation of higher spin currents essentially constrain the resulting three-point function to be of the form described in \cite{Giombi:2011ya, Maldacena:2012sf}.

It has been verified in \cite{Giombi:2009wh} in the limit of large $x_3$, and with $\lambda'$ taken to be equal to $\lambda$, that the result precisely agrees with the three-point function in the $O(N)$ vector model.\footnote{To fix the relative normalization, the computation of the same three-point function, with the spin $s$ and spin $0$ operators treated as sources while taking the boundary limiting value of the spin $s'$ field, with $s>s'$, was performed in \cite{Giombi:2009wh}. Precise agreement in the coefficient of the three point function as a function of $s$ and $s'$ was found.} Note that the parity odd contribution vanishes in this special limit. On the other hand, it is easy to see that the parity odd terms in (\ref{intres}) are generally non-vanishing. For instance, in the case $s'=1$, in the large $x_3$ limit, and generic $\lambda,\lambda'$,
%
the parity odd contribution is proportional to
\ie
\sin\theta_0 {(\lambda{\bf x}_{12}\sigma^z\lambda)^{s-1}(\lambda\lambda')(\lambda {\bf x}_{12}\sigma^z \lambda') \over (x_{12}^2)^{s+1} x_3^2}.
\fe
This is indeed the expected parity odd $s$-1-0 structure \cite{Giombi:2011rz} in the large $x_3$ limit.

So far we have restricted to the case $s>s'$. The case $s=s'$ is special. Naively, there is no contribution from $J^\Omega$. This is presumably an artifact due to the slightly singular nature of Vasiliev's system. We shall regularize the calculation by starting with the $s>s'$ case, analytically continue the result in the spin and take the $s\to s'$ limit. We obtain
\ie
&{\cal J}^+(\vec x,z) 
= -e^{i\theta_0}{2^{-2s-1} s} (2s)! {z^{2s+1}\over (x^2)^{2s+1}(\tilde x^2)^{2s+1}}(\lambda\sigma^z{\bf x} {\bf\tilde x} \sigma^z\lambda')^{2s}.
\fe
This contributes to the three point function $s$-$s$-$0$, 
\ie
& \int {dz d^3\vec x\over z^4} {z\over (\vec x-\vec x')^2 + z^2} z{\cal J}^+(\vec x,z)
\\
&=  -e^{i\theta_0}{2^{-2s-1} s(2s)!} \int {dz d^3\vec x} {z^{2s-1}\over [ (\vec x-\vec x')^2 + z^2] (x^2)^{2s+1}(\tilde x^2)^{2s+1}} (\lambda\sigma^z{\bf x} {\bf\tilde x}\sigma^z \lambda')^{2s}.
\fe
The parity even and odd terms are given by projecting the integrand onto odd and even functions in $z$, respectively. This gives precisely (B.5) of \cite{Maldacena:2012sf}, in the special case of spins $s$-$s$-$0$.

\subsection{Gauge function method} 

An important feature of Vasiliev's formulation of higher spin gauge theory is that the equation for the 1-form master field $W$ takes the form of a zero-curvature condition
\begin{equation}
d_xW+W*W=0\,.
\end{equation}
Locally, the solution always take a pure gauge form,
\begin{equation}
W=g^{-1}(x|Y,Z)*d_x g(x|Y,Z)\,,
\end{equation} 
where $g(x|Y,Z)$ is a ``gauge function" and $g^{-1}(x|Y,Z)$ its $*$-inverse.
Since Vasiliev's equations are gauge invariant, we can perform a formal gauge transformation by the function $g(x|Y,Z)$ to a new ``gauge", in which the master fields are denoted $W',B',S'$, with $W'= 0$ and 
\ie
&S(x|Y,Z) = g^{-1}(x|Y,Z)*d_Zg(x|Y,Z)+ g^{-1}(x|Y,Z)*S'(x|Y,Z)*g(x|Y,Z),\\
&B(x|Y,Z) = g^{-1}(x|Y,Z)*B'(x|Y,Z)*\pi(g(x|Y,Z)).
\label{g-transform}
\fe
We emphasize that this is not a true gauge transformation, in that the transformation by $g(x|Y,Z)$ does not respect the AdS boundary condition, and different solutions related by such a formal gauge transformation ought not to be thought of as physically equivalent, hence the quotation mark.
The equations of motion for the transformed fields are 
\ie
& d_x S'=0=d_x B',\\
& d_ZS'+S'*S' = B'*(e^{i\theta_0}K dz^2 + e^{-i\theta_0}\overline K d\bar z^2),\\
& d_Z B'+ S'*B' - B'*\pi(S')=0.
\label{Wzero-eq}
\fe
The first line simply implies that $S'(x|Y,Z)=S'(Y,Z)$ and $B'(x|Y,Z)=B'(Y,Z)$ are entirely independent of the space-time coordinates. Thus we are left with the task of solving the remaining two equations purely in the internal $(Y,Z)$ twistor space, which can be done perturbatively by simply integrating $Z$.  However, in order to extract the holographic correlation functions, we must go back to the physical spacetime by performing the gauge transformation (\ref{g-transform}) on the primed fields. Properly fixing the gauge function $g(x|Y,Z)$ is one of the main difficulties in applying this approach consistently.  

Clearly, one must require the function $g(x|Y,Z)$ to be such that at zeroth order in perturbation theory, it reproduces the $AdS_4$ vacuum solution for $W$. In other words, writing a perturbative expansion for the gauge function as 
\begin{equation}
g(x|Y,Z)=L(x|Y)+L(x|Y)*\epsilon^{(1)}(x|Y,Z)+\cdots,
\end{equation} 
the leading term $L(x|Y)$ should be related to the $AdS_4$ vacuum solution by
\ie
W_0(x|y,\bar y) = L^{-1}(x|y,\bar y)*d_xL(x|y,\bar y). 
\fe
We may write the solution of this equation as
\ie
L(x|Y) = {\bf P} \exp_*\left(-\int^{x_0}_x W_0^{\mu}(x'|Y) dx'_\mu\right)
\label{L-sol}
\fe
where ${\bf P} \exp_*$ stands for the path ordered $*$-exponential. The path goes from $x=(\vec x,z)$ to a fixed base point $x_0=(\vec x_0,z_0)$. The base point must lie in the bulk to ensure that the resulting $L(x|Y)$ is non-singular.
Without loss of generality, we may choose the base point to be $\vec x_0 =0$, $z_0=1$, in which case the explicit expression for the gauge function is
\ie
&L(x|Y) =\left[\det(\cosh M)\right]^{-{1\over 2}} \exp\left[{1\over 2}Y(\tanh M) Y\right],\\
&L^{-1}(x|Y) = \left[\det(\cosh M)\right]^{-{1\over 2}} \exp\left[-{1\over 2}Y(\tanh M) Y\right],
\fe
where $M$ is the $4\times 4$ matrix
\ie
M(x) = -{\ln z\over 4(z-1)} \left(\begin{matrix} \vec{x}\cdot \vec{\sigma}\sigma^z & \vec{x}\cdot \vec{\sigma}+(z-1)\sigma^z\\
\vec{x}\cdot \vec{\sigma}+(z-1)\sigma^z &\vec{x}\cdot \vec{\sigma}\sigma^z \end{matrix} \right).
\fe

In \cite{Giombi:2010vg} it was assumed that the first order correction to the gauge function $\epsilon^{(1)}(x|Y,Z)$ would not affect the result for the three point functions extracted from the boundary behavior of the second order $C$ master field. As we review below, this assumption allows for obtaining the correct result for the parity preserving three point functions in the type A and type B theories. However, the parity odd contributions in the general parity breaking Vasiliev's theory, which we derived in the physical spacetime approach for correlators of the type $s$-$s'$-$0$ ealier, could not be reproduced in this way. It appears that the ambiguity in properly fixing $\epsilon^{(1)}(x|Y,Z)$ should be closely related to the parity odd contributions, as will be discussed in more detail in section \ref{ambiguity} below. 

Let us now describe how the equations (\ref{Wzero-eq}) can be solved perturbatively. At linearized order, the equation for $B'$ simply implies that $B'^{(1)}(Y,Z)=B'^{(1)}(Y)$ is a function of $Y$ only. This function is fixed in terms of the physical bulk-to-boundary propagator $C^{(1)}(x|Y)$ given in (\ref{C-prop}) by performing the gauge transformation (\ref{g-transform}). We will describe this in detail in section \ref{twistor} below.

Once the linearized $B'$ field is known, we can then solve the linearized field $S'^{(1)}$ integrating 
\ie
d_Z S'^{(1)}= B'^{(1)}*( Kdz^2 + \overline Kd\bar z^2).
\fe
Here we have specialized to the type A theory, $\theta_0=0$. As already seen in the earlier analysis of linearized Vasiliev equations, the solution is
\ie
S'^{(1)} &=-z_\A dz^\A \int_0^1 dt\,t (B'^{(1)}*K)|_{z\to tz} + c.c.\\
&= -z_\A dz^\A \int_0^1 dt\,t B'^{(1)}(-tz,\bar y)K(t) + c.c.
\fe
where we have made the gauge choice $S'|_{Z=0}=0$. Note that this is not the same as the gauge condition $\left.S(x|Y,Z)\right|_{Z=0}=0$ \cite{Vasiliev:1995dn, Vasiliev:1999ba}, which should be imposed after transforming back to the physical gauge via the gauge function $g(x|Y,Z)$. We will discuss this issue later.
Carrying on to the second order, $B'^{(2)}$ and $S'^{(2)}$ can be solved from
\ie
& d_ZB'^{(2)} = -S'^{(1)}*B'^{(1)}+B'^{(1)}*\pi(S'^{(1)}),\\
& d_Z S'^{(2)} = -S'^{(1)}*S'^{(1)} + B'^{(2)}*(Kdz^2+\bar K d\bar z^2).
\fe
For computing tree-level three point functions, it suffices to solve $B'^{(2)}(Y,Z)$, which is explicitly given in terms of the linearized fields by
\ie
&B'^{(2)}(y,\bar y, z,\bar z) = -z^\A \int_0^1 dt \left[ S_\A'^{(1)}*B'^{(1)} - B'^{(1)}*\bar\pi(S_\A'^{(1)}) \right]_{z\to tz} + c.c.
\\
& = -2 \int d^4u d^4v e^{-uv+\bar u\bar v} B'^{(1)}(u,\bar u) B'^{(1)}(v,\bar v)
f(y,\bar y,z;U,V)
+c.c.
\label{Bprime2}
\fe
where, again, the appropriately rotated integration contour for $(u,v)$ and $(\bar u,\bar v)$ and the unconventionally normalized integration measure, coming from the integral representation of the star product, is understood. $U$ and $V$ denote $(u,\bar u)$ and $(v,\bar v)$ respectively. The function $f(y,\bar y,z;U,V)$ is obtained from a straightforward rewriting of the star product as (see \cite{Giombi:2010vg} for details)
\ie
f(y,\bar y,z;U,V) = \int_0^1 dt\int_0^\infty d\eta (zu) 
e^{(\eta u+tz)(y-v) + \bar y\bar u} \sinh\left( \bar y\bar v+ t zu \right).
\fe
Finally, we can recover the second order $B$ field in the physical spacetime by
\ie\label{btwo}
B^{(2)}(x|Y,Z)& = L^{-1}(x|Y) * B'^{(2)}(Y,Z) * \pi(L(x|Y))\\
&~~~- \epsilon^{(1)}(x|Y,Z)*B^{(1)}(x|Y) + B^{(1)}(x|Y)*\pi(\epsilon^{(1)}(x|Y,Z)),
\fe
and then take the AdS boundary limit $x= (\vec x,z\to 0)$ while restricting to $Z=0$ to extract the three point function (it suffices to further restrict to $\bar y=0$ or $y=0$, to extract the self-dual or the anti-self-dual part of the higher spin Weyl curvature).

\subsection{Bulk-to-boundary propagator and twistor transform}
\label{twistor}
The linearized field in the ``$W=0$ gauge" is related to the physical bulk-to-boundary propagator by 
\begin{equation}
B'^{(1)}(Y)=L(x|Y)*C^{(1)}(x|Y)*\pi(L^{-1}(x|Y))\,.
\end{equation}
By the definition, $L(x_0,Y)=1$, at the base point $x_0^\mu=(\vec x_0,z)$, see (\ref{L-sol}). So the linearized field in the $W=0$ gauge is simply
\ie
B'^{(1)}(Y) = C^{(1)}(x_0|Y).
\label{BprimeToC}
\fe
Explicitly, using eq. (\ref{C-prop}), choosing the base point $x_0=(\vec{0},z=1)$ and placing the higher spin boundary source at $\vec{x_1}$, we have  
\ie
B'^{(1)}_{(s)}(y,\bar y) = {(y(\vec{x}_1 \cdot \vec{\sigma}-\sigma^z ){\slash\!\!\!\varepsilon}\sigma^z(\vec{x}_1 \cdot \vec{\sigma}-\sigma^z) y)^s
\over (x_1^2+1)^{2s+1}}e^{-y(\sigma^z+2{\vec{x}_1 \cdot \vec{\sigma}-\sigma^z\over x_1^2+1})\bar y}+c.c.
\fe
As described earlier, it will be useful to express the null polarization vector as a spinor bilinear $({\slash\!\!\!\varepsilon}\sigma^z)_{\da\db} =\bar \lambda_\da \bar\lambda_\db$. In our conventions, we may also write $\bar\lambda = \sigma^z \lambda$. We can then construct a generating function for the boundary-to-bulk propagator associated with currents of all spins inserted at $\vec{x}_1$ and with polarization $\lambda$ 
\ie
B'^{(1)}(y,\bar y;\vec{x}_1,\lambda) 
&= {1\over x_1^2+1} e^{-y\left(\sigma^z+ 2 {\vec{x}_1 \cdot \vec{\sigma}-\sigma^z\over x_1^2+1}\right)\bar y}
\left\{ \exp\left[ - 2 y{\vec{x}_1 \cdot \vec{\sigma}-\sigma^z\over x_1^2+1}\bar \lambda \right]
+ \exp\left[ - 2 \bar y{\vec{x}_1 \cdot \vec{\sigma}-\sigma^z\over x_1^2+1} \lambda \right] \right\}.
\fe
It is convenient to define a Laplace transformed boundary-to-bulk propagator for $B'$ with respect to the polarization spinor $\lambda$,
\ie
&B_{tw}^{(1)}(y,\bar y;\mu) = {1\over 4} \int d^2\lambda e^{2\lambda\mu} B'^{(1)}(y,\bar y;\lambda)
 \\
 &=\delta\left( y + (\vec{x}_1 \cdot \vec{\sigma}\sigma^z-1)\mu  \right) 
e^{ -\mu ( \vec{x}_1 \cdot \vec{\sigma}-\sigma^z )\bar y } + \delta\left(\bar y-(\vec{x}_1 \cdot \vec{\sigma}-\sigma^z) \mu \right)
e^{ -y(\sigma^z\vec{x}_1 \cdot \vec{\sigma}+1)\mu }.
\fe
Let us further define
\ie
\bar\mu=-\sigma^z\mu,~~~~\chi = (\vec{x}_1 \cdot \vec{\sigma}-\sigma^z)\bar\mu,~~~~
\bar\chi =  (\vec{x}_1 \cdot \vec{\sigma}-\sigma^z)\mu,
\label{chi-def}
\fe
so that the Laplace transformed generating function for the propagator takes the remarkably simple form:
\ie\label{twprop}
&B_{tw}^{(1)}(y,\bar y;\chi,\bar\chi) = \delta(y- \chi)
e^{\bar\chi\bar y } + \delta\left(\bar y-\bar\chi \right)
e^{\chi y }.
\fe
We may regard $y,\bar y$ as independent holomorphic variables, and interpret the two terms in $B_{tw}^{(1)}$ as delta functions in the corresponding twistor space, where one of $y$ and $\bar y$ is Fourier/Laplace transformed.

\subsection{Three-point functions in the gauge function approach}
\subsubsection{A contour integral}

The first term on the RHS of (\ref{btwo}) is computed in \cite{Giombi:2010vg} by writing out the integral representation of the star products. After restricting to $Z=0$ and to $\bar y=0$ (the self-dual part of the Weyl curvature), and carefully taking the boundary limiting value, one indeed finds the expected falloff for the fields of each given spin in the Poincar\'e radial coordinate $z$ as $z\to 0$. Here we skip the tedious manipulation and give the result:
\ie\label{ftd}
&\lim_{z\to 0} z^{-1}B^{(2)}\!\left(\vec x=0,z\bigg|y_\A=z^{-{1\over 2}}\frac{w_\A}{2},\bar y_\da=0,Z=0\right)\\
&=-\int d^4u d^4v e^{uv-\bar u\bar v}B'^{(1)}(u,\bar u)B'^{(1)}(v,\bar v) \,(w\sigma^z\bar u)\left[ {e^{-w\sigma^z(\A_+-\bar u-\bar v)}\over (w\sigma^z\A_+)(\bar u\A_+)} -{e^{-w\sigma^z(\A_-+\bar u-\bar v)}\over (w\sigma^z\A_-)(\bar u\A_-)} \right].
\fe
The boundary point of interest, without loss of generality, is chosen to be the origin $\vec x=0$ (and $z=0$). Note that $\vec x=0$ is also the boundary coordinate of the base point, $(\vec x_0=0,z_0=1)$, which led to a simplification in (\ref{ftd}).
While the integration contour in $(u,\bar u,v,\bar v)$ was defined as the straight contours at 45 degree angle in the complex plane in the integral representation of the star product, and local deformations of the contour seemed unimportant, in (\ref{ftd}) the contour choice becomes important due to the poles of the integrand. The need for this choice of contour is presumably an artifact of the $W=0$ gauge, and should go away once we transform correctly back to the physical spacetime. The prescription of \cite{Giombi:2010vg} is such that the co-dimension 2 residues in $\A_\pm$ are picked up by the integration contour, giving the result
\ie
&\lim_{z\to 0} z^{-1}B^{(2)}\!\left(\vec x=0,z\bigg|y_\A=z^{-{1\over 2}}\frac{w_\A}{2},\bar y_\da=0,Z=0\right)\\
&=\int d^4u d^4v e^{uv-\bar u\bar v}B'^{(1)}(u,\bar u)B'^{(1)}(v,\bar v)\\
&~~~\times \left[ e^{-w\sigma^z(\bar u-\bar v)}\delta(\bar u-\bar v+\sigma^z(u+v))
+ e^{w\sigma^z (\bar u+\bar v)}\delta(\bar u+\bar v+\sigma^z(- u+ v)) \right].
\label{B2-final}
\fe

\subsubsection{A-type theory}

Earlier we wrote the linearized field $B'^{(1)}$ in the presence of a single boundary source current. To compute the three-point function, we can take two boundary sources, both contributing to $B'^{(1)}$, and take the cross term in (\ref{B2-final}). In other words, we should compute the contribution to (\ref{B2-final}) with $B'^{(1)}(U)$ and $B'^{(1)}(V)$ sourced by the two boundary currents respectively. Suppose the two boundary sources are located at $\vec x_1$ and $\vec x_2$, with their polarization specified by $\chi_{1,2}$ via the Laplace transform as in (\ref{chi-def}). Then, we have
\ie
&\lim_{z\to 0} z^{-1}B^{(2)}\left(\vec x=0,z \bigg|y_\A = z^{-{1\over 2}}(\lambda_3)_\A, \bar y_\da=0,Z=0;\chi_1,\chi_2\right)\\
&= 2 \cosh\left({\chi_1\chi_2+\bar\chi_1\bar\chi_2}\right) \left[ e^{2\lambda_3(\chi_1+\chi_2)} \delta(\chi_1+\chi_2+\sigma^z(\bar\chi_1+\bar\chi_2)) + \delta(\chi_1+\chi_2+\sigma^z(\bar\chi_1+\bar\chi_2)+2\lambda_3) \right]
\\
&~~~+(\chi_1\to -\chi_1)+(\chi_2\to -\chi_2)+(\chi_1\to -\chi_1,\chi_2\to -\chi_2).
\fe
Expressed in terms of $\mu_1, \mu_2$ (recall the definitions (\ref{chi-def})), this is
\ie\label{resa}
&\lim_{z\to 0} z^{-1}B^{(2)}(\vec x=0,z|z^{-{1\over 2}}y,\bar y=Z=0;\chi_1,\chi_2)\\
&={1\over 2} \cosh\left({2\mu_1\sigma^z{\bf x_{12}}\mu_2} \right) \left[ e^{2\lambda_3 (\mu_1+\mu_2)} \delta({\bf x_1}\mu_1+{\bf x_2}\mu_2) + \delta({\bf x_1}\mu_1+{\bf x_2}\mu_2+\sigma^z \lambda_3) \right]
\\
&~~~+(\mu_1\to -\mu_1)+(\mu_2\to -\mu_2)+(\mu_1\to -\mu_1,\mu_2\to -\mu_2).
\fe
Here we used the notation ${\bf x}_i\equiv \vec x_i\cdot\vec \sigma$.
After Laplace transforming back from $\mu_1,\mu_2$ to the polarization spinors $\lambda_1,\lambda_2$ of the two source currents, and shifting the third operator from the origin $\vec x=0$ to $\vec x = \vec x_3$, we obtain the generating function of all three point functions,
\begin{eqnarray}
\label{ttm}
&&{4\over |x_{12}||x_{23}| |x_{31}|}
\cosh\left( {1\over 2}\lambda_1\sigma^z {\bf \check x_{12}x_{23}\check x_{13}}\lambda_1
+{1\over 2}\lambda_2\sigma^z{\bf \check x_{23} x_{31} \check x_{21}}\lambda_2 +{1\over 2} \lambda_3\sigma^z{\bf \check x_{31}x_{12}\check x_{32}}\lambda_3 \right)\cr
&&~~~\times \cosh\left(\lambda_1 \sigma^z{{\bf \check x_{12}}}\lambda_2\right)\cosh\left( \lambda_1\sigma^z {{\bf\check x_{13}}}\lambda_3 \right)\cosh\left( \lambda_2\sigma^z {{\bf\check x_{23}}}\lambda_3 \right).
\end{eqnarray}
Recall the notation ${\bf \check x_i} = \vec x_i\cdot\vec \sigma/x_i^2$.
The correlator of three higher spin currents $\langle J_{s_1}(x_1;\lambda_1) J_{s_2}(x_2;\lambda_2) J_{s_3}(x_3;\lambda_3)\rangle$ is read off from the terms of order $\lambda_1^{2s_1} \lambda_2^{2s_2} \lambda_3^{2s_3}$. It can be checked that this indeed is a generating function for three point functions of higher spin currents made out of bilinears of a free massless scalar field in three dimensions. At this point, the coupling constant
$g$ of Vasiliev theory must be put in by hand, as the overall coefficient of (\ref{ttm}), if the two-point functions are normalized to be independent of $g$. Comparing with $O(N)$ vector model, $g$ is identified with $1/\sqrt{N}$ up a numerical coefficient that was determined in \cite{Giombi:2009wh}.

\subsubsection{B-type theory}

Repeating the calculation in B type theory, we again start from (\ref{B2-final}) but with a slightly different expression for the Laplace transformed boundary-to-bulk propagator (from (\ref{C-prop}) with $\theta_0=\pi/2$)
\ie\label{twb}
&B_{tw;B}^{(1)}(y,\bar y;\chi,\bar\chi) = i\delta(y- \chi)
e^{\bar\chi\bar y } - i\delta\left(\bar y-\bar\chi \right)
e^{\chi y }.
\fe
Inserting this into (\ref{B2-final}), performing the integral in twistor space and Fourier transforming back to $\lambda_1,\lambda_2$
\begin{eqnarray}
\label{jjj-F}
&&{4\over |x_{12}||x_{23}| |x_{31}|}
\sinh\left( {1\over 2}\lambda_1\sigma^z {\bf \check x_{12}x_{23}\check x_{13}}\lambda_1
+{1\over 2}\lambda_2\sigma^z{\bf \check x_{23} x_{31} \check x_{21}}\lambda_2 +{1\over 2} \lambda_3\sigma^z{\bf \check x_{31}x_{12}\check x_{32}}\lambda_3 \right)\cr
&&~~~\times \sinh\left(\lambda_1 \sigma^z{{\bf \check x_{12}}}\lambda_2\right) \sinh\left( \lambda_1\sigma^z {{\bf\check x_{13}}}\lambda_3 \right) \sinh\left( \lambda_2\sigma^z {{\bf\check x_{23}}}\lambda_3 \right).
\end{eqnarray}
Indeed, this is a generating function for three point functions of currents of nonzero spins in the free fermion theory. 

Note that the $\Delta=2$ scalar operator must be treated separately, as it is not included in the generating function for the boundary-to-bulk propagator (\ref{twb}). Here we give some details of the computation of the three-point function involving a scalar in the $B$ type theory (previously unpublished though mentioned in \cite{Giombi:2011rz}). We will take the outgoing field to be a spin $s_3$ field inserted at the origin, and the two sources to be a spin $s_1$ current at $\vec{x}_1$ with polarization $\lambda_1$ and a $\Delta=2$ scalar inserted at $\vec{x}_2$. The bulk to boundary propagator for the spin $s_1$ field is given by (\ref{twb}), while for the $\Delta=2$ scalar we need to start from (\ref{C-prop-Del2}) and evaluate it at the base point of the gauge function, as in (\ref{BprimeToC}). This yields, for the scalar inserted at the boundary point $\vec{x}_2$
\ie
B'^{(1)}_{\Delta=2}(Y) 
&= {1\over (x_2^2+1)^2} \left[1-y(\sigma^z + 2 {\vec x_2\cdot \vec\sigma-\sigma^z\over x_2^2+1})\bar y\right] e^{-y(\sigma^z + 2 {\vec x_2\cdot \vec\sigma-\sigma^z\over x_2^2+1}) \bar y}
\\
&= \left(1+\xi\partial_\xi|_{\xi=1}\right) {e^{-\xi y(\sigma^z + 2 {\vec x_2\cdot \vec\sigma-\sigma^z\over x_2^2+1}) \bar y} \over (x_2^2+1)^2}
\\
& = {1+\xi\partial_\xi|_{\xi=1}\over x_2^2+1} \int d^2\mu \delta(y-\chi_2) e^{\xi \bar\chi_2 \bar y},
\fe
where $\chi_2$ and $\bar\chi_2$ are defined as $\chi_2 = (\vec x_2\cdot\vec \sigma-\sigma^z)\bar\mu,\bar\chi_2 = (\vec x_2\cdot\vec \sigma - \sigma^z)\mu, \bar\mu = -\sigma^z\mu$.

The boundary expectation value of the outcoming $B$ field is obtained by inserting the propagators for the spin $s_1$ field and the $\Delta=2$ scalar into (\ref{B2-final}). We need to calculate
\ie
&{1+\xi\partial_\xi|_{\xi=1}\over x_2^2+1} \int d^4u d^4v e^{uv-\bar u\bar v} \left[ i\delta(u-\chi_1)e^{\bar\chi_1\bar u} -i \delta(\bar u-\bar\chi_1) e^{\chi_1 u} \right] \delta(v-\chi_2) e^{\xi \bar\chi_2\bar  v}
\\
&~~~\times \left[ e^{-2y\sigma^z(\bar u-\bar v)} \delta(\bar u-\bar v + \sigma^z(u+v))
+ e^{2y\sigma^z(\bar u+\bar v)} \delta(\bar u+\bar v + \sigma^z(-u+v)) \right]
\\
&=i{1+\xi\partial_\xi|_{\xi=1}\over x_2^2+1} \left\{ e^{\chi_1\chi_2+\xi\bar\chi_1\bar\chi_2 }\left[
e^{2y(\chi_1+\chi_2)} \delta(\chi_1+\chi_2+\sigma^z(\bar\chi_1+\xi\bar\chi_2)) 
+e^{2y(\chi_1-\chi_2)}\delta(\chi_1-\chi_2+\sigma^z(\bar\chi_1-\xi\bar\chi_2))
\right]
\right.
\\
&~~~~\left.
- 2\cosh \left(\chi_1\chi_2 + \xi \bar\chi_1\bar\chi_2\right) \delta(2y+\chi_1-\chi_2+\sigma^z(\bar\chi_1-\xi\bar\chi_2)) \right\}.
\fe
Implicitly, the truncation condition on Vasiliev's master fields implies a projection on the boundary-to-bulk propagators, retaining integer spin fields. We are thus instructed to take the even part in $\chi_2$, sum over the contribution with the roles of $\chi_1$ and $\chi_2$ interchanged, and write the resulting contribution to the boundary limiting value of $B$ as
\ie
&2i{1+\xi\partial_\xi|_{\xi=1}\over x_2^2+1} \cosh \left(\chi_1\chi_2 + \xi \bar\chi_1\bar\chi_2\right)\left\{ e^{2y(\chi_1+\chi_2)} \delta(\chi_1+\chi_2+\sigma^z(\bar\chi_1+\xi\bar\chi_2)) 
\right.
\\
&\left.
- \delta(2y+\chi_1+\chi_2+\sigma^z(\bar\chi_1+\xi\bar\chi_2)) +(\chi_1\to -\chi_1)+ (\chi_2\to -\chi_2) + (\chi_1\to -\chi_1,\chi_2\to -\chi_2) \right\}.
\fe
In terms of $\mu_1,\mu_2$, we can write it as
\ie
&{i\over 2}{1+\partial_\eta|_{\eta=0}\over x_2^2+1} \cosh \left[2\mu_1\sigma^z{\bf x_{12}}\mu_2 + \eta\mu_1({\bf x_1}-\sigma^z)({\bf x_2}-\sigma^z)\mu_2\right]
\\
&\times \left\{ e^{2y(\mu_1+\mu_2+{\eta\over2}(1-\sigma^z{\bf x_2})\mu_2)} \delta( {\bf x_1}\mu_1+{\bf x_2}\mu_2 + {\eta\over 2}({\bf x_2}-\sigma^z)\mu_2) -  \delta(\sigma^z y+{\bf x_1}\mu_1+{\bf x_2}\mu_2 + {\eta\over 2}({\bf x_2}-\sigma^z)\mu_2)
\right.
\\
&\left.
 +(\mu_1\to -\mu_1)+ (\mu_2\to -\mu_2) + (\mu_1\to -\mu_1,\mu_2\to -\mu_2) \right\}
\fe
Next, we integrate $\mu_2$ to recover the contribution from the $\Delta=2$ scalar operator at $\vec x_2$, yielding
\ie
&2i{1+\partial_\eta|_{\eta=0}\over x_2^2+1} \left\{ {1-\eta\over x_2^2} \cosh \left({2\over x_2^2}\mu_1\sigma^z{\bf x_{12}x_2x_1}\mu_1 \right)
 \right.
\\
&\left.\times \left[
\cosh\left(  {2\over x_2^2}\mu_1 {\bf x_{12}x_2} y +  \eta(1+x_2^2) \mu_1 {\sigma^z{\bf x_1}\over x_2^2} y \right)-\cosh\left( {2\over x_2^2}\mu_1 {\bf x_{12}x_2} y - \eta (1+x_2^2)\mu_1{\sigma^z{\bf x_1}\over x_2^2} y \right)\right]  \right\}\\
&=\frac{4i}{x_2^2} \left(\mu_1{\sigma^z{\bf x_1}\over x_2^2} y \right)\cosh \left({2\over x_2^2}\mu_1\sigma^z{\bf x_{12}x_2x_1}\mu_1 \right)\sinh\left( {2\over x_2^2}\mu_1 {\bf x_{12}x_2} y \right)
\fe 
Laplace transforming $\mu_1$ back to $\lambda_1$, setting $y=\lambda_3$, and shifting the origin to $\vec x_3$, we obtain
\ie
{2i\over |x_{13}| |x_{23}| |x_{12}|}\left( \lambda_1{{\bf \check x_{12}\check x_{32}}} \lambda_3 \right) \cosh\left( {1\over 2} \lambda_1\sigma^z {\bf \check x_{12}x_{23}\check x_{13}}\lambda_1
+ {1\over 2}\lambda_3\sigma^z{\bf \check x_{31}x_{12}\check x_{32}}\lambda_3
\right)\,\sinh\left( \lambda_1\sigma^z {\bf \check x_{13}}\lambda_3 \right)
\fe
This is indeed the generating function for the three point functions $\langle J_{s_1}(\vec{x}_1;\lambda_1)O_{\Delta=2}(\vec{x}_2)J_{s_3}(\vec{x}_3;\lambda_3)\rangle$
in the free fermion theory (see \cite{Giombi:2011rz}). 

One may analogously compute the three point function with two $\Delta=2$ scalars and one higher spin current. Taking the two scalars to be the sources, we start from 
\ie
&{(1+\xi_1\partial_{\xi_1})(1+\xi_2\partial_{\xi_2})|_{\xi_1=\xi_2=1}\over (x_1^2+1)(x_2^2+1)} \int d^4u d^4v e^{uv-\bar u\bar v} \delta(u-\chi_1) e^{\xi_1 \bar\chi_1\bar  u} \delta(v-\chi_2) e^{\xi_2 \bar\chi_2\bar  v}
\\
&~~~\times \left[ e^{-2y\sigma^z(\bar u-\bar v)} \delta(\bar u-\bar v + \sigma^z(u+v))
+ e^{2y\sigma^z(\bar u+\bar v)} \delta(\bar u+\bar v + \sigma^z(-u+v)) \right]
\fe 
and proceed similarly as above. After the twistor space and $\mu_1,\mu_2$ integrals are performed, one finds that the final result is proportional to 
\ie 
\frac{\lambda_3\sigma^z{\bf \check x_{31}x_{12} \check x_{32}}\lambda_3}{2|x_{12}|^3 |x_{23}||x_{13}|} 
\sinh\left({1\over 2} \lambda_3\sigma^z{\bf\check x_{31}x_{12}\check x_{32}}\lambda_3\right)\,.
\fe 
This indeed generates the correct three point functions $\langle O_{\Delta=2}(\vec{x}_1)O_{\Delta=2}(\vec{x}_2)J_{s_3}(\vec{x}_3;\lambda_3)\rangle$.

\subsection{Gauge ambiguity}
\label{ambiguity}

Thus far we have ignored the ``gauge correction" term involving $\epsilon^{(1)}(x|Y,Z)$ in extracting the boundary correlator from (\ref{btwo}). Recall in the analysis of linearized equations that in order to identify the components of $B$ master field restricted to $Z=0$ with the higher spin Weyl curvature, we worked in the gauge $S|_{Z=0}=0$. This gauge condition should be maintained in order to extract the correct physical higher spin fields from Vasiliev's master fields. After transforming $S', B'$ to the physical spacetime, we have, at linearized order,
\ie
&W^{(1)}(x|Y,Z) = D_0 \epsilon^{(1)}(x|Y,Z) ,\\
&S^{(1)}(x|Y,Z) = L^{-1}(x|Y)*S'^{(1)}(Y,Z)*L(x|Y) + d_Z \epsilon^{(1)}(x|Y,Z).
\fe
While $S'|_{Z=0}=0$, $(L^{-1}*S'*L)|_{Z=0}$ is generally nonzero, and thus a nonzero first order gauge correction by $\epsilon^{(1)}$ is generally needed. The condition $S^{(1)}|_{Z=0}$ by itself, however, does not fix the part of $\epsilon^{(1)}$ that is quadratic and higher order in $Z$. This ambiguity does not affect the analysis of linearized equations, but generally matter at higher order in perturbation theory. In the physical spacetime calculation of correlators, we have used a solution of $S^{(1)}$ with the property that $S^{(1)}_\A$ is proportional to $z_\A$ while $S^{(1)}_\da$ is proportional to $\bar z_\da$. This suggests a gauge condition
\ie
z^\A S_\A + \bar z^\da S_\da = 0.
\fe
When combined with $S|_{Z=0}=0$, it fixes the $Z$-dependence of $\epsilon(x|Y,Z)$ entirely.

There is a superficial argument that suggests $\epsilon^{(1)}$ would not affect the boundary values of the second order fields, in some cases.
Near the AdS boundary $z\to 0$ ($z$ is the Poincar\'e radial coordinate, not to be confused with the twistor variables $z_\A, \bar z_\da$), the spin-$s$ component of $\Omega(\vec x,z|Y)$ falls off like $z^s$, whereas
the spin-$s$ component of $C(\vec x,z|Y)$ falls off like $z^{s+1}$. It is then natural to impose the $z^s$ fall-off condition on the spin-$s$ component of the gauge function $\epsilon^{(1)}(x|Y,Z)$. So one may expect the ``gauge correction" in (\ref{btwo}) to fall off like $z^{s_1+s_2+1}$, which would not affect the leading boundary behavior of the spin-$s_3$ component of $B^{(2)}$, provided $s_3<s_1+s_2$. Given three operators of spins $s_1, s_2, s_3$ (not all zero), we can always choose two spins, say $s_1, s_2$, with $s_3<s_1+s_2$, and compute the three point function regarding $J^{(s_1)}$ and $J^{(s_2)}$ as sources. Indeed, the physical spacetime computation in \cite{Giombi:2009wh} was successfully carried out in this case, for the parity invariant theory. However, it is possible that inverse powers of Poincar\'e radial coordinate $z$ are introduced in taking the star product of boundary-to-bulk propagators of master fields. When this happens, the argument given above breaks down. We believe that in going from the ``$W=0$ gauge" to the physical spacetime, a correction due to $\epsilon^{(1)}$ is generally required, for the parity violating theory with parity breaking phase $\theta_0$. In fact, the $s$-$s'$-$0$ correlator computed in section \ref{sspz}, which successfully produced the parity odd contributions and received contribution only from $J^\Omega$, would come entirely from the gauge correction! A proper treatment of the gauge correction in the gauge function approach remains to be done, and we hope to revisit it in the near future.

\section{Summary and open questions}

We have reviewed the construction of Vasiliev's pure higher spin gauge theories in $AdS_4$, and their conjectured dualities with free and critical large $N$ vector models, as well as with Chern-Simons vector models in the parity violating generalization. We formulated Vasiliev perturbation theory around the $AdS_4$ vacuum, the goal being computing boundary correlators of higher spin currents. This is achieved explicitly in the case of tree level three-point functions, by treating two of the boundary currents as sources, and the three-point function as the boundary limiting value of the field dual to the third boundary operator. The computation in the most straightforward ``physical spacetime approach", starting with the boundary-to-bulk propagators of the master fields and solving the full spacetime dependence of the master fields at the quadratic order, is particularly simply when one of the three boundary operators is scalar. We gave an explicit integral expression for the $s$-$s'$-$0$ correlator, and showed in some examples that it agrees with the explicitly known structures constrained by higher spin symmetry broken by $1/N$ effects. The computation we presented here refines that of \cite{Giombi:2009wh}, in that we keep track of the full position and polarization dependence of these correlators, in the Vasiliev theory with parity breaking phase $\theta_0$, which in particular allows us to extract the parity odd contributions to the three-point functions. The dependence of the three-point function on $\theta_0$ agrees with the general form constrained by slightly broken higher spin symmetry \cite{Maldacena:2012sf}.

We then turned to a different method of computing correlators, by first working in the ``$W=0$ gauge", where the spacetime dependence of the master fields are formally gauged away and consequently Vasiliev's equations can be solved order by order simply by integrating in the $Z$-twistor variables. One then transform back to the physical gauge to extract boundary correlators \cite{Giombi:2010vg}. With some hand waiving, we were able to produce the complete three-point functions of currents of all spins that match precisely with the result of free and critical $O(N)$ or $U(N)$ vector models. There are however a few potentially important caviats that have not been understood properly. The first is that we have not carefully fixed the ambiguity due to a possible gauge transformation by a first order gauge parameter, in computing the second order fields sourced by two boundary sources. The second is that we encountered a singular integral representation of the star product, in transforming the second order field back to the physical gauge and taking its boundary limiting value. A contour prescription was given to regularize the integral in twistor space. This contour prescription is a priori unjustified, although the result agrees with the computation in the physical spacetime approach for $s$-$s'$-$0$ correlators. We believe that this singular nature of the $W=0$ gauge computation is tied to the gauge ambiguity, and the latter needs to be understood in order to generalize the $W=0$ gauge computation to the parity violating case and higher point correlators.

So what comprises a proof of the higher spin/vector model duality? As shown in \cite{Giombi:2011ya}, if the duality holds for one set of $AdS$ boundary conditions, say between minimal A-type Vasiliev theory with $\Delta=1$ boundary condition and the free $O(N)$ vector model, then the duality between the same bulk theory with an alternative $\Delta=2$ boundary condition and the critical $O(N)$ model also holds, at least to all order in perturbation theory in $1/N$. It was shown in \cite{Maldacena:2011jn} that a conformal field theory with {\it exact} higher spin symmetry must be a free theory. If the spectrum of single trace operators, dual to single particle states in the bulk, are entirely given by the conserved higher spin currents, then it must be a vector model. The remaining question is then whether the bulk Vasiliev theory indeed has a set of boundary conditions that allow for the higher spin symmetry to be preserved, and that the higher spin symmetry is preserved upon quantization. This is not entirely obvious: for instance the theory with a generic parity breaking phase $\theta_0$ does not admit any boundary conditions that preserve the higher spin symmetries \cite{Chang:2012kt}. It was nonetheless shown in \cite{Chang:2012kt} that the $\Delta=1$ boundary condition indeed preserve higher spin symmetries of the $A$-type theory, while the $\Delta=2$ boundary condition breaks them. 

So far there has been little understanding in the quantization of Vasiliev theory, for several reasons. Firstly, an explicit action which at quadratic level reduces to the ordinary free field actions is not available at present. Interesting proposals for an action for Vasiliev's theory were studied in \cite{Doroud:2011xs} and \cite{Boulanger:2011dd, Boulanger:2012bj} (see also \cite{Vasiliev:1988sa} for earlier related suggestions). However, they do not appear to reproduce the standard Fronsdal action at the free field level. In the absence of a conventional action, it is unclear how to fix to a covariant gauge and couple the ghosts to the higher spin fields, thus preventing a covariant quantization. It may be possible to quantize the theory in a gauge where the ghosts decouple, such as the lightcone gauge, where some of the symmetries are no longer manifest. There is potentially also the issue of regularization, since it is not clear how to deform the $AdS_4$ Vasiliev theory to $d=4-\epsilon$ dimensions while preserving higher spin gauge symmetry. In the case of parity invariant theories with higher spin symmetry preserving boundary condition, the dual CFT is free which implies that in the bulk computation of correlators of boundary currents, the tree-level result is exact. In this case, all higher loop corrections in the bulk must exactly cancel. In the parity violating theories, however, the quantum corrections in the bulk are presumably nonzero and is expected to reproduce the $1/N$ expansion of correlators of higher spin currents in Chern-Simons vector models, which is nontrivial.

The conjectured dualities between parity violating Vasiliev theory and Chern-Simons vector models suggests the intriguing possibility of ``three-dimensional bosonization" \cite{Giombi:2011kc, Maldacena:2012sf, OferUpcoming}. Namely, the Vasiliev theory with parity breaking phase $\theta_0$ and $\Delta=1$ boundary condition is expected to be dual to CS-scalar vector model with 't Hooft coupling $\lambda=N/k$ and the identification $\theta_0={\pi\over 2}\lambda$, while the Vasiliev theory with $\Delta=2$ boundary condition and parity breaking phase $\theta_0={\pi\over 2}(1-\lambda)$ is expected to be dual to CS-fermion vector model with 't Hooft coupling $\lambda$. Assuming that the higher order coefficients $\theta_2, \theta_4$ etc. in the function $\theta(X) = \theta_0 + \theta_2 X^2 + \theta_4 X^4+\cdots$ are absent, the duality would imply that the Chern-Simone theory coupled to critical boson, i.e. the critical point of CS-scalar vector model deformed by the square of the dimension 1 scalar operator, is the same CFT as the CS-fermion vector model, under the identification $\lambda\to 1-\lambda$ and appropriate rescaling in $N$. And vice versa, it would also imply a strong-weak duality between the CS-critical-fermion vector model and the CS-scalar vector model. However, the computation of free energy of the CS-vector models on the plane at finite temperature, by solving the theory in lightcone gauge at large $N$ using Schwinger-Dyson equations, seems to be in conflict with such a bosonization duality \cite{Giombi:2011kc, Jain:2012qi}. A potential resolution to this puzzle is that $\theta_2, \theta_4$ etc. are nonzero and play a role in the duality between parity violating Vasiliev theory and Chern-Simons vector models, and their values for the dual of CS-scalar and CS-fermion theories do not agree under $\lambda\to 1-\lambda$. If so, the three-dimensional bosonization would not hold, despite the agreement of three-point functions at large $N$ \cite{OferUpcoming}. Even if this is the case, the situation is not entirely satisfactory as we do not understand the role of $\theta_2, \theta_4$ etc. in the consistency of the bulk theory itself. A priori $\theta_2, \theta_4$ etc. appear to be arbitrary and cannot be removed by field redefinition, and are consistent with the higher spin symmetry which is ``slightly broken" by boundary conditions. If all possible values of $\theta_n$'s give rise to consistent higher spin gauge theories in $AdS_4$, they should all be dual to some parity non-invariant vector models, while there appears to be no candidate of the dual CFT besides Chern-Simons vector models. Perhaps a sharp test would be the explicit computation of the contribution of say $\theta_2$ to the 5-point function. We hope to investigate this in the near future.

The dualities between supersymmetric Vasiliev theory, possibly with nontrivial Chan-Paton factors, and supersymmetric Chern-Simons vector models, are explored in \cite{Chang:2012kt}. It is satisfactory that one can identify the precise boundary conditions of the bulk theory that preserve ${\cal N}=0,1,2,3,$ and $4$ or $6$ supersymmetries, that are dual to Chern-Simons vector models with the same matter content but differ by double-trace and triple-trace deformations, as well as possibly gauging a flavor group with another Chern-Simons gauge field. In fact, had one not known the existence of ${\cal N}=4$ or ${\cal N}=6$ Chern-Simons theories, one would discover them by seeing that there are boundary conditions of supersymmetric Vasiliev theory that preserve these numbers of supersymmetries. The duality indicates that Vasiliev theory with Chan-Paton factors is generally dual to the quiver Chern-Simons-matter theory, viewed as a vector model with its flavor group gauged. It suggests a concrete embedding of a supersymmetric Vasiliev theory with ${\cal N}=6$ boundary condition into type IIA string theory, namely, the dual of ABJ model, type IIA string theory on $AdS_4\times \mathbb{CP}^3$ with flat $B$-field, should be equivalent in the small radius limit to the Vasiliev theory in $AdS_4$.

While the equivalence between the above mentioned supersymmetric Vasiliev theory and type IIA closed string field theory in $AdS_4\times \mathbb{CP}^3$ is expected because they have the same CFT dual, we are not suggesting that the two bulk theories have the same classical equations of motion. This is because a general single closed string state is mapped to a multi-particle state of higher spin fields under the correspondence, while the single higher spin particle should be mapped to the leading Regge trajectory among the closed string states.

There is, however, a more straightforward way to engineer Chern-Simons vector models in string theory, starting with the dual of $U(N)_k\times U(M)_{-k}$ ABJ theory, namely type IIA strings in $AdS_4\times \mathbb{CP}^3$ with flat $B$-field, add $N_f$ D6-branes wrapped on $AdS_4\times \mathbb{RP}^3$ which introduce $N_f$ hypermultiplets in the fundamental representation of the $U(N)$, and then take the ``minimal radius limit" $M=0$. In this limit, the closed string sector should somehow become topological (dual to the pure $U(N)$ Chern-Simons; see also \cite{Banerjee:2012gh}) while the open strings on the D6-branes are the only propagating degrees of freedom in $AdS_4$. The dual CFT, which is the ${\cal N}=3$ $U(N)$ Chern-Simons vector model with $N_f$ hypermultiplets, is also expected to be dual to a supersymmetric Vasiliev theory with ${\cal N}=3$ boundary condition, described in \cite{Chang:2012kt}. In this case, we anticipate that the open string field equations on the D6-branes in the minimal radius limit should literally reduce to that of Vasiliev's system. To show this directly in the bulk requires a concrete understanding of the very stringy limit of the open string field theory, which should be possible in the pure spinor formalism \cite{Berkovits:2007zk, Berkovits:2008qc, Berkovits:2008ga, Mazzucato:2011jt}.

We have focused on perturbation theory, and have not discussed solutions of Vasiliev theory that represent large, finite deformation away from the $AdS_4$ vacuum. The simplest such solution was found in \cite{Sezgin:2005pv}, where a nontrivial profile for the bulk scalar (and only the scalar) is turned on. Its Euclidean continuation describes the Euclidean $AdS_4$ solution with $S^3$ conformal boundary, with a finite scalar field turned on that is invariant under the $SO(4)$ isometry that rotates the $S^3$. This solution of A-type Vasiliev theory is presumably dual to the mass deformed $O(N)$ vector model on $S^3$. There are also nontrivial exact solutions that appear to respect the ordinary $AdS_4$ boundary condition \cite{Didenko:2009td, Iazeolla:2011cb}. In some gauge, the graviton sector of the solution of \cite{Didenko:2009td} appears to be identical to that of the global AdS-Schwarzschild black hole, despite that the solution appears to be also extremal and can be embedded as a BPS solution in the supersymmetric higher spin theory. A careful analysis of the asymptotic charges of the solution remains to be done. In fact, the existence of black hole solution in Vasiliev theory in global $AdS_4$ would seem to conflict with the dual vector model on $S^2$, as the gauge singlet constraint on the Hilbert space of the latter prevents order $N$ free energy at temperature order 1 in units of the radius of the sphere. The physical meaning of the solutions of \cite{Didenko:2009td} (and generalized in \cite{Iazeolla:2011cb}) and their role in the higher spin/vector model duality are thus far unclear.

A rich story that has been left out entirely in this review is the $AdS_3/CFT_2$ version of the higher spin/vector model duality. The two-dimensional analog of the singlet $O(N)$ or $U(N)$ vector model, as observed by Gaberdiel and Gopakumar \cite{Gaberdiel:2012uj}, is the $W_N$ minimal model. A complication of the two-dimensional story is that the operators of the CFT are in correspondence with states of the CFT on the circle, and that the circle is not simply connected, which requires including twisted sector states in gauging the theory. In the discussion of three-dimensional vector models, we have mostly restricted ourselves to the study of correlation functions of gauge invariant local operators on $\mathbb{R}^3$. If one is to study the (Chern-Simons) vector model on topologically nontrivial 3-manifolds \cite{Shenker:2011zf, Banerjee:2012gh}, the singlet condition amounts to integrating over flat connections, which can be singular in the limit where the Chern-Simons level $k$ is taken to infinity. In fact, a finite Chern-Simons level is required to make sense of the Hilbert space of the singlet vector model on a Riemann surface of genus greater than zero, and one finds a large density of states in the large $k$ limit \cite{Banerjee:2012gh}. To make sense of the duality between $AdS_4$ Vasiliev theory and the three-dimensional Chern-Simons vector model beyond correlators on $\mathbb{R}^3$ or $\mathbb{S}^3$ presumably requires an extension of Vasiliev theory to include some sort of topological sector (this is strongly suggested by the duality between supersymmetric Vasiliev theory and type IIA string theory with D6-branes on $AdS_4\times\mathbb{CP}^3$, as mentioned earlier). This should also be relevant to explain the fact that the free energy of the Chern-Simons vector model on $\mathbb{S}^3$ is expected to contain a term proportional to $N^2$ \cite{Klebanov:2011gs}, essentially coming from the pure Chern-Simons contribution, while the bulk coupling constant in Vasiliev's theory should scale as $g_{\rm bulk}^{-2}\sim N$.  In higher than two-dimensions, a conformal field theory is often regarded as being characterized entirely by its spectrum of gauge invariant local operators and their correlation functions. In this sense, the duality makes sense and is consistent without the need of consideration of boundaries of nontrivial topology. One does not have this option in two-dimensions, however. The inclusion of twisted sector states is enforced by modular invariance, and in the case of $W_N$ minimal model, this essentially leads to a large density of lower dimension operators in the large $N$ limit (the ``light states"). The large $N$ factorization holds provided that one identifies a large number of the light states as ``single-trace" operators, dual to one-particle states in the bulk $AdS_3$. Thus, the holographic dual of the $W_N$ minimal model should be a higher spin gauge theory coupled to an infinite tower of matter fields. Vasiliev's system in $AdS_3$, unlike the four-dimensional one, is not a pure higher spin gauge theory. Its spectrum consists of a tower of gauge fields of spin $s=2,3,4,\cdots$, coupled to a single complex massive scalar field. It was conjectured that Vasiliev's system in $AdS_3$ by itself is in fact dual to a subsector of $W_N$ minimal model, {\it perturbatively} in $1/N$. This subsector consists a subset of $W_N$ primaries whose OPEs close at infinite $N$, and whose correlators on the plane or the sphere makes sense, and are expected to match with those of the $AdS_3$ Vasiliev theory perturbatively, order by order in $1/N$. This is perhaps the more precise $AdS_3/CFT_2$ analog of the $AdS_4/CFT_3$ higher spin/vector model dualities. On the other hand, it suggests that Vasiliev theory is non-perturbatively incomplete, and that the complete non-perturbative quantum higher spin theory requires adding {\it perturbative} states.

To summarize, the higher spin/vector model dualities provide a rich class of examples of holographic dualities with and without supersymmetry. They allow us to explore perturbative holography in great detail, and compare the bulk and boundary theories directly, order by order in perturbation theory (with respect to the gauge coupling in the bulk and with respect to $1/N$ in the boundary). In some cases it may be viewed as a limit of string field theory. While much work remains just to understand the perturbative duality (such as quantum corrections in the bulk), the non-perturbative aspects are much more mysterious and intriguing. It remains to be seen what kind of lessons on quantum gravity can be drawn by exploring this corner of AdS/CFT correspondence.

\section*{Acknowledgements}

First and foremost, we would like to thank our collaborators C.-M. Chang, S. Minwalla, S. Prakash, T. Sharma, S. Shenker, S. Trivedi and S. Wadia. We are grateful to O. Aharony, V.E. Didenko, N. Doroud, A. Dymarsky, M. Gaberdiel, C. Iazeolla, D. Jafferis, I. Klebanov, J. Maldacena, R. Myers, S. Rey, E. Sezgin, A. Strominger, P. Sundell, M.A. Vasiliev and A. Zhiboedov for many useful discussions. We would like to thank the Aspen Center for Physics and the NSF Grant 1066293 for hospitality during completion of this work. S.G. also thanks the Simons Center for Geometry and Physics for hospitality during the 2012 Summer Simons Workshop in Mathematics and Physics.  S.G. is supported by Perimeter Institute for Theoretical Physics. Research at Perimeter Institute is supported by the Government of Canada through Industry Canada and by the Province of Ontario through
the Ministry of Research $\&$ Innovation. The work of X.Y. is supported in part by the Fundamental Laws Initiative Fund at Harvard University, and by NSF Award PHY-0847457.

\bibliographystyle{utphys}
\bibliography{HSreview}

\providecommand{\href}[2]{#2}\begingroup\raggedright\begin{thebibliography}{10}

\bibitem{Klebanov:2002ja}
I.~Klebanov and A.~Polyakov, ``{AdS dual of the critical O(N) vector model},''
  \href{http://dx.doi.org/10.1016/S0370-2693(02)02980-5}{{\em Phys.Lett.}
  {\bfseries B550} (2002) 213--219},
\href{http://arxiv.org/abs/hep-th/0210114}{{\ttfamily arXiv:hep-th/0210114
  [hep-th]}}.

\bibitem{Sezgin:2003pt}
E.~Sezgin and P.~Sundell, ``{Holography in 4D (super) higher spin theories and
  a test via cubic scalar couplings},''
  \href{http://dx.doi.org/10.1088/1126-6708/2005/07/044}{{\em JHEP} {\bfseries
  07} (2005) 044},
\href{http://arxiv.org/abs/hep-th/0305040}{{\ttfamily arXiv:hep-th/0305040}}.

\bibitem{Sundborg:1999ue}
B.~Sundborg, ``{The Hagedorn transition, deconfinement and N=4 SYM theory},''
  \href{http://dx.doi.org/10.1016/S0550-3213(00)00044-4}{{\em Nucl.Phys.}
  {\bfseries B573} (2000) 349--363},
\href{http://arxiv.org/abs/hep-th/9908001}{{\ttfamily arXiv:hep-th/9908001
  [hep-th]}}.

\bibitem{HaggiMani:2000ru}
P.~Haggi-Mani and B.~Sundborg, ``{Free large N supersymmetric Yang-Mills theory
  as a string theory},'' {\em JHEP} {\bfseries 0004} (2000) 031,
\href{http://arxiv.org/abs/hep-th/0002189}{{\ttfamily arXiv:hep-th/0002189
  [hep-th]}}.

\bibitem{Konstein:2000bi}
S.~Konstein, M.~Vasiliev, and V.~Zaikin, ``{Conformal higher spin currents in
  any dimension and AdS / CFT correspondence},'' {\em JHEP} {\bfseries 0012}
  (2000) 018,
\href{http://arxiv.org/abs/hep-th/0010239}{{\ttfamily arXiv:hep-th/0010239
  [hep-th]}}.

\bibitem{Shaynkman:2001ip}
O.~Shaynkman and M.~Vasiliev, ``{Higher spin conformal symmetry for matter
  fields in (2+1)-dimensions},''
  \href{http://dx.doi.org/10.1023/A:1012399417069}{{\em Theor.Math.Phys.}
  {\bfseries 128} (2001) 1155--1168},
\href{http://arxiv.org/abs/hep-th/0103208}{{\ttfamily arXiv:hep-th/0103208
  [hep-th]}}.

\bibitem{Sezgin:2001zs}
E.~Sezgin and P.~Sundell, ``{Doubletons and 5-D higher spin gauge theory},''
  {\em JHEP} {\bfseries 0109} (2001) 036,
\href{http://arxiv.org/abs/hep-th/0105001}{{\ttfamily arXiv:hep-th/0105001
  [hep-th]}}.

\bibitem{Vasiliev:2001zy}
M.~Vasiliev, ``{Conformal higher spin symmetries of 4-d massless
  supermultiplets and osp(L,2M) invariant equations in generalized
  (super)space},'' \href{http://dx.doi.org/10.1103/PhysRevD.66.066006}{{\em
  Phys.Rev.} {\bfseries D66} (2002) 066006},
\href{http://arxiv.org/abs/hep-th/0106149}{{\ttfamily arXiv:hep-th/0106149
  [hep-th]}}.

\bibitem{Mikhailov:2002bp}
A.~Mikhailov, ``{Notes on higher spin symmetries},''
\href{http://arxiv.org/abs/hep-th/0201019}{{\ttfamily arXiv:hep-th/0201019
  [hep-th]}}.

\bibitem{Sezgin:2002rt}
E.~Sezgin and P.~Sundell, ``{Massless higher spins and holography},''
  \href{http://dx.doi.org/10.1016/S0550-3213(02)00739-3}{{\em Nucl.Phys.}
  {\bfseries B644} (2002) 303--370},
\href{http://arxiv.org/abs/hep-th/0205131}{{\ttfamily arXiv:hep-th/0205131
  [hep-th]}}.

\bibitem{Giombi:2011kc}
S.~Giombi, S.~Minwalla, S.~Prakash, S.~P. Trivedi, S.~R. Wadia, {\em et~al.},
  ``{Chern-Simons Theory with Vector Fermion Matter},''
\href{http://arxiv.org/abs/1110.4386}{{\ttfamily arXiv:1110.4386 [hep-th]}}.

\bibitem{Giombi:2009wh}
S.~Giombi and X.~Yin, ``{Higher Spin Gauge Theory and Holography: The
  Three-Point Functions},''
  \href{http://dx.doi.org/10.1007/JHEP09(2010)115}{{\em JHEP} {\bfseries 1009}
  (2010) 115},
\href{http://arxiv.org/abs/0912.3462}{{\ttfamily arXiv:0912.3462 [hep-th]}}.

\bibitem{Giombi:2010vg}
S.~Giombi and X.~Yin, ``{Higher Spins in AdS and Twistorial Holography},''
  \href{http://dx.doi.org/10.1007/JHEP04(2011)086}{{\em JHEP} {\bfseries 1104}
  (2011) 086},
\href{http://arxiv.org/abs/1004.3736}{{\ttfamily arXiv:1004.3736 [hep-th]}}.

\bibitem{Chang:2012kt}
C.-M. Chang, S.~Minwalla, T.~Sharma, and X.~Yin, ``{ABJ Triality: from Higher
  Spin Fields to Strings},''
\href{http://arxiv.org/abs/1207.4485}{{\ttfamily arXiv:1207.4485 [hep-th]}}.

\bibitem{Giombi:2011rz}
S.~Giombi, S.~Prakash, and X.~Yin, ``{A Note on CFT Correlators in Three
  Dimensions},''
\href{http://arxiv.org/abs/1104.4317}{{\ttfamily arXiv:1104.4317 [hep-th]}}.

\bibitem{Costa:2011mg}
M.~S. Costa, J.~Penedones, D.~Poland, and S.~Rychkov, ``{Spinning Conformal
  Correlators},'' \href{http://dx.doi.org/10.1007/JHEP11(2011)071}{{\em JHEP}
  {\bfseries 1111} (2011) 071},
\href{http://arxiv.org/abs/1107.3554}{{\ttfamily arXiv:1107.3554 [hep-th]}}.

\bibitem{Zhiboedov:2012bm}
A.~Zhiboedov, ``{A note on three-point functions of conserved currents},''
\href{http://arxiv.org/abs/1206.6370}{{\ttfamily arXiv:1206.6370 [hep-th]}}.

\bibitem{Maldacena:2011jn}
J.~Maldacena and A.~Zhiboedov, ``{Constraining Conformal Field Theories with A
  Higher Spin Symmetry},''
\href{http://arxiv.org/abs/1112.1016}{{\ttfamily arXiv:1112.1016 [hep-th]}}.

\bibitem{Maldacena:2012sf}
J.~Maldacena and A.~Zhiboedov, ``{Constraining conformal field theories with a
  slightly broken higher spin symmetry},''
\href{http://arxiv.org/abs/1204.3882}{{\ttfamily arXiv:1204.3882 [hep-th]}}.

\bibitem{OferUpcoming}
O.~Aharony, G.~Gur-Ari, and R.~Yacoby, ``{Correlation Functions of Large N
  Chern-Simons-Matter Theories and Bosonization in Three Dimensions},''
\href{http://arxiv.org/abs/1207.4593}{{\ttfamily arXiv:1207.4593 [hep-th]}}.

\bibitem{Jain:2012qi}
S.~Jain, S.~P. Trivedi, S.~R. Wadia, and S.~Yokoyama, ``{Supersymmetric
  Chern-Simons Theories with Vector Matter},''
\href{http://arxiv.org/abs/1207.4750}{{\ttfamily arXiv:1207.4750 [hep-th]}}.

\bibitem{Douglas:2010rc}
M.~R. Douglas, L.~Mazzucato, and S.~S. Razamat, ``{Holographic dual of free
  field theory},'' \href{http://dx.doi.org/10.1103/PhysRevD.83.071701}{{\em
  Phys.Rev.} {\bfseries D83} (2011) 071701},
\href{http://arxiv.org/abs/1011.4926}{{\ttfamily arXiv:1011.4926 [hep-th]}}.

\bibitem{Koch:2010cy}
R.~d.~M. Koch, A.~Jevicki, K.~Jin, and J.~P. Rodrigues, ``{$AdS_4/CFT_3$
  Construction from Collective Fields},''
  \href{http://dx.doi.org/10.1103/PhysRevD.83.025006}{{\em Phys.Rev.}
  {\bfseries D83} (2011) 025006},
\href{http://arxiv.org/abs/1008.0633}{{\ttfamily arXiv:1008.0633 [hep-th]}}.

\bibitem{Jevicki:2011ss}
A.~Jevicki, K.~Jin, and Q.~Ye, ``{Collective Dipole Model of AdS/CFT and Higher
  Spin Gravity},'' \href{http://dx.doi.org/10.1088/1751-8113/44/46/465402}{{\em
  J.Phys.A} {\bfseries A44} (2011) 465402},
\href{http://arxiv.org/abs/1106.3983}{{\ttfamily arXiv:1106.3983 [hep-th]}}.

\bibitem{Jevicki:2011aa}
A.~Jevicki, K.~Jin, and Q.~Ye, ``{Bi-local Model of AdS/CFT and Higher Spin
  Gravity},''
\href{http://arxiv.org/abs/1112.2656}{{\ttfamily arXiv:1112.2656 [hep-th]}}.

\bibitem{deMelloKoch:2012vc}
R.~de~Mello~Koch, A.~Jevicki, K.~Jin, J.~P. Rodrigues, and Q.~Ye, ``{S=1 in
  O(N)/HS duality},''
\href{http://arxiv.org/abs/1205.4117}{{\ttfamily arXiv:1205.4117 [hep-th]}}.

\bibitem{Das:2003vw}
S.~R. Das and A.~Jevicki, ``{Large N collective fields and holography},''
  \href{http://dx.doi.org/10.1103/PhysRevD.68.044011}{{\em Phys.Rev.}
  {\bfseries D68} (2003) 044011},
\href{http://arxiv.org/abs/hep-th/0304093}{{\ttfamily arXiv:hep-th/0304093
  [hep-th]}}.

\bibitem{Anninos:2011ui}
D.~Anninos, T.~Hartman, and A.~Strominger, ``{Higher Spin Realization of the
  dS/CFT Correspondence},''
\href{http://arxiv.org/abs/1108.5735}{{\ttfamily arXiv:1108.5735 [hep-th]}}.

\bibitem{Ng:2012xp}
G.~S. Ng and A.~Strominger, ``{State/Operator Correspondence in Higher-Spin
  dS/CFT},''
\href{http://arxiv.org/abs/1204.1057}{{\ttfamily arXiv:1204.1057 [hep-th]}}.

\bibitem{Das:2012dt}
D.~Das, S.~R. Das, A.~Jevicki, and Q.~Ye, ``{Bi-local Construction of Sp(2N)/dS
  Higher Spin Correspondence},''
\href{http://arxiv.org/abs/1205.5776}{{\ttfamily arXiv:1205.5776 [hep-th]}}.

\bibitem{Gaberdiel:2012uj}
M.~R. Gaberdiel and R.~Gopakumar, ``{Minimal Model Holography},''
\href{http://arxiv.org/abs/1207.6697}{{\ttfamily arXiv:1207.6697 [hep-th]}}.

\bibitem{Didenko:2012vh}
V.~Didenko and E.~Skvortsov, ``{Towards higher-spin holography in ambient space
  of any dimension},''
\href{http://arxiv.org/abs/1207.6786}{{\ttfamily arXiv:1207.6786 [hep-th]}}.

\bibitem{Vasiliev:1995dn}
M.~A. Vasiliev, ``{Higher spin gauge theories in four-dimensions,
  three-dimensions, and two-dimensions},''
  \href{http://dx.doi.org/10.1142/S0218271896000473}{{\em Int.J.Mod.Phys.}
  {\bfseries D5} (1996) 763--797},
\href{http://arxiv.org/abs/hep-th/9611024}{{\ttfamily arXiv:hep-th/9611024
  [hep-th]}}.

\bibitem{Vasiliev:1999ba}
M.~A. Vasiliev, ``{Higher spin gauge theories: Star product and AdS space},''
\href{http://arxiv.org/abs/hep-th/9910096}{{\ttfamily arXiv:hep-th/9910096
  [hep-th]}}.

\bibitem{Bekaert:2005vh}
X.~Bekaert, S.~Cnockaert, C.~Iazeolla, and M.~Vasiliev, ``{Nonlinear higher
  spin theories in various dimensions},''
\href{http://arxiv.org/abs/hep-th/0503128}{{\ttfamily arXiv:hep-th/0503128
  [hep-th]}}.

\bibitem{Iazeolla:2008bp}
C.~Iazeolla, ``{On the Algebraic Structure of Higher-Spin Field Equations and
  New Exact Solutions},''
\href{http://arxiv.org/abs/0807.0406}{{\ttfamily arXiv:0807.0406 [hep-th]}}.

\bibitem{Engquist:2002vr}
J.~Engquist, E.~Sezgin, and P.~Sundell, ``{On N=1, N=2, N=4 higher spin gauge
  theories in four-dimensions},''
  \href{http://dx.doi.org/10.1088/0264-9381/19/23/316}{{\em Class.Quant.Grav.}
  {\bfseries 19} (2002) 6175--6196},
\href{http://arxiv.org/abs/hep-th/0207101}{{\ttfamily arXiv:hep-th/0207101
  [hep-th]}}.

\bibitem{Hartman:2006dy}
T.~Hartman and L.~Rastelli, ``{Double-trace deformations, mixed boundary
  conditions and functional determinants in AdS/CFT},''
  \href{http://dx.doi.org/10.1088/1126-6708/2008/01/019}{{\em JHEP} {\bfseries
  0801} (2008) 019},
\href{http://arxiv.org/abs/hep-th/0602106}{{\ttfamily arXiv:hep-th/0602106
  [hep-th]}}.

\bibitem{Giombi:2011ya}
S.~Giombi and X.~Yin, ``{On Higher Spin Gauge Theory and the Critical O(N)
  Model},''
\href{http://arxiv.org/abs/1105.4011}{{\ttfamily arXiv:1105.4011 [hep-th]}}.

\bibitem{Girardello:2002pp}
L.~Girardello, M.~Porrati, and A.~Zaffaroni, ``{3-D interacting CFTs and
  generalized Higgs phenomenon in higher spin theories on AdS},''
  \href{http://dx.doi.org/10.1016/S0370-2693(03)00492-1}{{\em Phys.Lett.}
  {\bfseries B561} (2003) 289--293},
\href{http://arxiv.org/abs/hep-th/0212181}{{\ttfamily arXiv:hep-th/0212181
  [hep-th]}}.

\bibitem{Gaiotto:2007qi}
D.~Gaiotto and X.~Yin, ``{Notes on superconformal Chern-Simons-Matter
  theories},'' \href{http://dx.doi.org/10.1088/1126-6708/2007/08/056}{{\em
  JHEP} {\bfseries 0708} (2007) 056},
\href{http://arxiv.org/abs/0704.3740}{{\ttfamily arXiv:0704.3740 [hep-th]}}.

\bibitem{Aharony:2011jz}
O.~Aharony, G.~Gur-Ari, and R.~Yacoby, ``{d=3 Bosonic Vector Models Coupled to
  Chern-Simons Gauge Theories},''
  \href{http://dx.doi.org/10.1007/JHEP03(2012)037}{{\em JHEP} {\bfseries 1203}
  (2012) 037},
\href{http://arxiv.org/abs/1110.4382}{{\ttfamily arXiv:1110.4382 [hep-th]}}.

\bibitem{Sezgin:2002ru}
E.~Sezgin and P.~Sundell, ``{Analysis of higher spin field equations in
  four-dimensions},'' {\em JHEP} {\bfseries 0207} (2002) 055,
\href{http://arxiv.org/abs/hep-th/0205132}{{\ttfamily arXiv:hep-th/0205132
  [hep-th]}}.

\bibitem{Witten:1998qj}
E.~Witten, ``{Anti-de Sitter space and holography},'' {\em
  Adv.Theor.Math.Phys.} {\bfseries 2} (1998) 253--291,
\href{http://arxiv.org/abs/hep-th/9802150}{{\ttfamily arXiv:hep-th/9802150
  [hep-th]}}.

\bibitem{Vasiliev:1990bu}
M.~A. Vasiliev, ``{Algebraic aspects of the higher spin problem},''
\href{http://dx.doi.org/10.1016/0370-2693(91)90867-P}{{\em Phys.Lett.}
  {\bfseries B257} (1991) 111--118}.

\bibitem{Bolotin:1999fa}
K.~Bolotin and M.~A. Vasiliev, ``{Star product and massless free field dynamics
  in AdS(4)},'' \href{http://dx.doi.org/10.1016/S0370-2693(00)00307-5}{{\em
  Phys.Lett.} {\bfseries B479} (2000) 421--428},
\href{http://arxiv.org/abs/hep-th/0001031}{{\ttfamily arXiv:hep-th/0001031
  [hep-th]}}.

\bibitem{Sezgin:2005pv}
E.~Sezgin and P.~Sundell, ``{An Exact solution of 4-D higher-spin gauge
  theory},'' \href{http://dx.doi.org/10.1016/j.nuclphysb.2006.06.038}{{\em
  Nucl.Phys.} {\bfseries B762} (2007) 1--37},
\href{http://arxiv.org/abs/hep-th/0508158}{{\ttfamily arXiv:hep-th/0508158
  [hep-th]}}.

\bibitem{Iazeolla:2007wt}
C.~Iazeolla, E.~Sezgin, and P.~Sundell, ``{Real forms of complex higher spin
  field equations and new exact solutions},''
  \href{http://dx.doi.org/10.1016/j.nuclphysb.2007.08.002}{{\em Nucl.Phys.}
  {\bfseries B791} (2008) 231--264},
\href{http://arxiv.org/abs/0706.2983}{{\ttfamily arXiv:0706.2983 [hep-th]}}.

\bibitem{Iazeolla:2011cb}
C.~Iazeolla and P.~Sundell, ``{Families of exact solutions to Vasiliev's 4D
  equations with spherical, cylindrical and biaxial symmetry},''
  \href{http://dx.doi.org/10.1007/JHEP12(2011)084}{{\em JHEP} {\bfseries 1112}
  (2011) 084},
\href{http://arxiv.org/abs/1107.1217}{{\ttfamily arXiv:1107.1217 [hep-th]}}.

\bibitem{Doroud:2011xs}
N.~Doroud and L.~Smolin, ``{An Action for higher spin gauge theory in four
  dimensions},''
\href{http://arxiv.org/abs/1102.3297}{{\ttfamily arXiv:1102.3297 [hep-th]}}.

\bibitem{Boulanger:2011dd}
N.~Boulanger and P.~Sundell, ``{An action principle for Vasiliev's
  four-dimensional higher-spin gravity},''
  \href{http://dx.doi.org/10.1088/1751-8113/44/49/495402}{{\em J.Phys.A}
  {\bfseries A44} (2011) 495402},
\href{http://arxiv.org/abs/1102.2219}{{\ttfamily arXiv:1102.2219 [hep-th]}}.

\bibitem{Boulanger:2012bj}
N.~Boulanger, N.~Colombo, and P.~Sundell, ``{A minimal BV action for Vasiliev's
  four-dimensional higher spin gravity},''
\href{http://arxiv.org/abs/1205.3339}{{\ttfamily arXiv:1205.3339 [hep-th]}}.

\bibitem{Vasiliev:1988sa}
M.~A. Vasiliev, ``{CONSISTENT EQUATIONS FOR INTERACTING MASSLESS FIELDS OF ALL
  SPINS IN THE FIRST ORDER IN CURVATURES},''
\href{http://dx.doi.org/10.1016/0003-4916(89)90261-3}{{\em Annals Phys.}
  {\bfseries 190} (1989) 59--106}.

\bibitem{Banerjee:2012gh}
S.~Banerjee, S.~Hellerman, J.~Maltz, and S.~H. Shenker, ``{Light States in
  Chern-Simons Theory Coupled to Fundamental Matter},''
\href{http://arxiv.org/abs/1207.4195}{{\ttfamily arXiv:1207.4195 [hep-th]}}.

\bibitem{Berkovits:2007zk}
N.~Berkovits, ``{A New Limit of the AdS(5) x S**5 Sigma Model},''
  \href{http://dx.doi.org/10.1088/1126-6708/2007/08/011}{{\em JHEP} {\bfseries
  0708} (2007) 011},
\href{http://arxiv.org/abs/hep-th/0703282}{{\ttfamily arXiv:hep-th/0703282
  [hep-th]}}.

\bibitem{Berkovits:2008qc}
N.~Berkovits, ``{Perturbative Super-Yang-Mills from the Topological AdS(5) x
  S**5 Sigma Model},''
  \href{http://dx.doi.org/10.1088/1126-6708/2008/09/088}{{\em JHEP} {\bfseries
  0809} (2008) 088},
\href{http://arxiv.org/abs/0806.1960}{{\ttfamily arXiv:0806.1960 [hep-th]}}.

\bibitem{Berkovits:2008ga}
N.~Berkovits, ``{Simplifying and Extending the AdS(5) x S**5 Pure Spinor
  Formalism},'' \href{http://dx.doi.org/10.1088/1126-6708/2009/09/051}{{\em
  JHEP} {\bfseries 0909} (2009) 051},
\href{http://arxiv.org/abs/0812.5074}{{\ttfamily arXiv:0812.5074 [hep-th]}}.

\bibitem{Mazzucato:2011jt}
L.~Mazzucato, ``{Superstrings in AdS},''
\href{http://arxiv.org/abs/1104.2604}{{\ttfamily arXiv:1104.2604 [hep-th]}}.

\bibitem{Didenko:2009td}
V.~Didenko and M.~Vasiliev, ``{Static BPS black hole in 4d higher-spin gauge
  theory},'' \href{http://dx.doi.org/10.1016/j.physletb.2009.11.023}{{\em
  Phys.Lett.} {\bfseries B682} (2009) 305--315},
\href{http://arxiv.org/abs/0906.3898}{{\ttfamily arXiv:0906.3898 [hep-th]}}.

\bibitem{Shenker:2011zf}
S.~H. Shenker and X.~Yin, ``{Vector Models in the Singlet Sector at Finite
  Temperature},''
\href{http://arxiv.org/abs/1109.3519}{{\ttfamily arXiv:1109.3519 [hep-th]}}.

\bibitem{Klebanov:2011gs}
I.~R. Klebanov, S.~S. Pufu, and B.~R. Safdi, ``{F-Theorem without
  Supersymmetry},'' \href{http://dx.doi.org/10.1007/JHEP10(2011)038}{{\em JHEP}
  {\bfseries 1110} (2011) 038},
\href{http://arxiv.org/abs/1105.4598}{{\ttfamily arXiv:1105.4598 [hep-th]}}.

\end{thebibliography}\endgroup
\end{document}